\documentclass[aps,prd,secnumarabic,amssymb,nobibnotes, amsmath,twocolumn,nofootinbib,superscriptaddress,floatfix,preprintnumbers]{revtex4-1}

\usepackage{graphicx,color}% Include figure files
\usepackage{dcolumn}% Align table columns on decimal point
\usepackage{bm}% bold math
\usepackage{lineno, soul}
\usepackage[normalem]{ulem}
\usepackage{hyperref}% add hypertext capabilities
\usepackage{amsmath}
\usepackage{mathtools}
\usepackage{multirow}
\usepackage{textgreek}
\usepackage{afterpage}
\usepackage{aas_macros}
\usepackage{xcolor}
\usepackage{comment}
\usepackage{rotating}

\definecolor{darkcyan}{rgb}{0.0, 0.55, 0.55}
\definecolor{purple2}{rgb}{0.5, 0.0, 0.5}

\newcommand{\tas}{Tibet AS$\gamma$}
\newcommand{\pg}{$p-\gamma$}
\newcommand{\gag}{g_{a\gamma\gamma}}

\newcommand{\Fermi}{{\it Fermi}}
\newcommand{\diff}{{\mathrm{d}}}

\def \aap  {A\&A}

\def \aj  {AJ}
\def \apj  {ApJ}
\def \apjs  {ApJS}
\def \apjl  {ApJL}

\def \jcap  {JCAP}
\def \prd {Phy. Rev. D}

\def \mnras {MNRAS}
\def \nat {Nature}

\def \prl {PRL}

\begin{document}

\preprint{LAPTH-026/22}

\title{First constraints on axion-like particles from Galactic sub-PeV gamma rays
}

\author{C.~Eckner}
\email{eckner@lapth.cnrs.fr}
\affiliation{LAPTh, USMB, CNRS,  F-74940 Annecy, France}
\author{F.~Calore}
\email{calore@lapth.cnrs.fr}
\affiliation{LAPTh, USMB, CNRS,  F-74940 Annecy, France}

\begin{abstract}
Experimental refinements and technical innovations in the field of extensive air shower telescopes have enabled measurements of Galactic cosmic-ray interactions in the sub-PeV 
%\new{\sout{(100 TeV to 1 PeV)}} 
range, providing new avenues for the search for new physics and dark matter.
For the first time, we exploit sub-PeV (1 TeV -- 1 PeV) observations of Galactic diffuse gamma rays by
HAWC and Tibet AS$\gamma$ %\sout{Tibet AS$\gamma$ and HAWC}}
to search for an axion-like-particle (ALP) induced gamma-ray signal 
directly linked to the origin of the IceCube extragalactic high-energy neutrino flux.
Indeed, the production of high-energy neutrinos in extragalactic sources implies the concomitant production of gamma rays at comparable energies. Within the magnetic field of the neutrino emitting sources, gamma rays may efficiently convert into ALPs, escape their host galaxy un-attenuated, propagate through intergalactic space, and reconvert into gamma rays in the magnetic field of the Milky Way. Such a scenario creates an all-sky diffuse high-energy gamma-ray signal in the sub-PeV range.
Accounting for the guaranteed Galactic astrophysical gamma-ray contributions from cosmic-ray interactions with gas and radiation and from sub-threshold sources, 
we set competitive upper limits on the photon-ALP coupling constant $g_{a\gamma\gamma}$. 
We find $g_{a\gamma\gamma} < 2.1\times10^{-11}$ GeV$^{-1}$ for ALP masses $m_a \leq 2\times10^{-7}$ eV at a 95\% confidence level. Our results are comparable to previous limits on ALPs derived from the TeV gamma-ray domain and progressively close the mass gap towards ADMX limits. The code and data to reproduce the results of this study are available on GitHub \url{https://github.com/ceckner/subPeVALPs}.
\end{abstract}

\maketitle

%%%%%%%%%%%%%%%%%%%%%%%%%%%%%%%%%%%%%%
\section{Introduction}
\label{sec:intro}
%%%%%%%%%%%%%%%%%%%%%%%%%%%%%%%%%%%%%%
The advent of large-field-of-view ground-based telescopes,
such as e.g.~Tibet AS$\gamma$~\cite{Amenomori:2019rjd} and LHAASO~\cite{LHAASO:2019qtb},
has recently opened a new astrophysical window on the very-high-energy gamma-ray sky
by measuring, for the first time, the diffuse Galactic emission at sub-PeV energies
and superseding previous upper limits from CASA-MIA~\cite{Borione:1997fy} and IceTop~\cite{IceCube:2019hmk}.
This is complemented at lower energies, $E_\gamma \gtrsim 1$ TeV,
by ARGO-YBJ~\cite{ARGO-YBJ:2015cpa} and, more recently, by HAWC~\cite{HAWC:2021bvb} measurements of the diffuse Galactic emission 
over extended regions of the sky.

Without considering exotic physics phenomena --as discussed in the main 
body of this work--,
these observations are supposed to be purely of Galactic origin,
from cosmic-ray interactions with gas and radiation fields, as
well as from the cumulative contribution of faint, i.e.~{\it unresolved}, Galactic sources. 
The ``standard'' extragalactic emission from active galactic nuclei, normal galaxies, etc.~is 
indeed believed to be negligible at sub-PeV energies because of 
absorption (and subsequent cascades) of sub-PeV gamma rays on the cosmic microwave background (CMB) and the extragalactic background light (EBL), whose main components are the cosmic infrared background (CIB; originating from re-radiation of light absorbed by dust particles) and the cosmic optical background (COB; radiation created by stars and galaxies) both being at about $10\%$ of the intensity of the CMB \cite{Driver:2021sne}. The CMB and EBL together render the universe almost opaque to sub-PeV gamma rays. For example, the mean free path of 1 PeV gamma rays is limited to tens of kpc, i.e.~if observed, such emission can only be related to Galactic astrophysical processes. At 10 TeV the mean free path may be as large as tens of Mpc allowing observers on Earth the study of the closest blazars \cite{Kohri:2012tq, Ruffini:2015oha}.
Sub-PeV gamma-ray observations have been used to constrain the population of cosmic-ray electrons~\cite{2021ApJ...919...93F},
protons and nuclei~\cite{2021PhRvD.104d3010K} in the Galactic disk, 
also in synergy with the astrophysical neutrino flux measurement~\cite{2021ApJ...914L...7L}, starting to provide a unique insight onto the nature of cosmic-ray interactions~\cite{Luque:2022buq}.
%
%E.g.~\cite{2021ApJ...919...93F} shows that ``data imply either  an additional component in the cosmic-ray nucleon spectrum or  contribution from discrete sources, including PeVatrons such as superbubbles and hypernova remnants, and PeV electron accelerators''. 
%
%Other possible explanations: \cite{2021ApJ...914L...7L}; \cite{2021PhRvD.104d3010K};  \cite{2021A&A...653L...4N}

Eventually, exotic processes which produce sub-PeV photons can also supply part of the observed emission.
First implications of \tas~data have been derived, for example, for heavy decaying
dark matter~\cite{2021PhRvD.104b1301E,2021arXiv210505680N}.
We focus here on axion-like particles (ALPs), elusive pseudo-scalar particles often 
predicted in multiple extensions of the Standard Model of particle physics, e.g.~\cite{Peccei:1977hh, Weinberg:1977ma, Wilczek:1977pj, Kim:1979if, Shifman:1979if, Zhitnitsky:1980tq, Georgi:1981pu, Dine:1981rt, Witten:1984dg, Raffelt:1996wa, Conlon:2006tq, Arvanitaki:2009fg, Choi:2009jt, Cicoli:2012sz, Dias:2014osa, Ringwald:2014vqa}, 
which can also represent viable dark matter candidates~\cite{Preskill:1982cy, Abbott:1982af, Dine:1982ah, Arias:2012az} in some portions of their 
parameter space~\cite{Wantz:2009it, Sikivie:2006ni, Hiramatsu:2012gg, AxionLimits}.
%
%The goal of the present work is to assess what is the parameter space of axion-like particles that can be constrained with the sub-PeV diffuse signal.
We consider ALPs $a$ minimally coupled with the photon via the Lagrangian \cite{Raffelt:1987im}
\begin{equation}
\label{eq:ALP_lagrangian}
\mathcal{L}_{a\gamma}=-\frac{1}{4} \gag F_{\mu\nu}\tilde{F}^{\mu\nu}a=\gag {\bf E}\cdot{\bf B}\,a~\mathrm{,}
\end{equation}
thus inducing a conversion of ALPs into photons (propagating as an electromagnetic wave ${\bf E}$ with wave vector ${\bf k}$ linked to its energy via $E = |{\bf k}|$) and vice versa in the presence of an external magnetic field ${\bf B}$, the so-called Primakoff process~\cite{Raffelt:1985nk}.
Various processes of ALP production and photon conversion 
can be tested with high-energy gamma-ray astrophysics, and typically 
allow us to set constraints on ALPs in the mass of $10^{-11} - 10^{-6}$ eV.
We refer the reader to the summary plots of Ref.~\cite{AxionLimits} and references therein.

This conversion mechanism implies that, among other effects, a fraction of photons produced by 
high-energy astrophysical sources, instead of being absorbed through interactions
with internal radiation fields or the CMB and EBL, may convert into ALPs which then travel unimpeded
and reconvert into photons in the Milky Way's magnetic field. Such a way to alter the
transparency of the universe has been discussed thoroughly in the past, also highlighting 
some possible anomalies in TeV data~\cite{DeAngelis:2008sk,Horns:2012fx}.
Therefore, the existence of ALPs 
enlarges the gamma-ray horizon, and
makes the detections of extragalactic
very high-energy photons smoking gun signatures of their nature.

At the highest energies, the production of photons in astrophysical sources is accompanied by
the production of a neutrino flux, either through $p-p$ or $p-\gamma$ interactions. 
In particular, the sub-PeV neutrino flux measured by IceCube is believed to originate from
extragalactic sources and cosmic-ray interactions therein~\cite{Gaggero:2015xza, Feyereisen:2017fnk, Denton:2017csz, ANTARES:2017nlh, ANTARES:2018nyb, IceCube:2019lzm}. 
%According to the consensus theory for the production of very-high-energy cosmic rays, the neutrino production mechanisms do likewise produce an associated gamma-ray flux. 
Multi-messenger analyses constrain the contribution of different source classes to the 
neutrino diffuse flux, in particular the ones from $p-p$ interactions, which is severely challenged by \Fermi-LAT gamma-ray data in the GeV domain~\cite{Murase:2013rfa, Tamborra:2014xia, Bechtol:2015uqb, 2015arXiv151101530K, Ando:2015bva}. 

The goal of the present work is to assess what is the ALPs parameter space that can be constrained with the sub-PeV (1 TeV -- 1 PeV) diffuse signal, exploiting the synergy with the astrophysical neutrino flux.
In particular, we will follow the model of Ref.~\cite{2015arXiv151101530K, 2017arXiv171201839V} and quantify the cumulative ALP-induced gamma-ray flux from a population of extragalactic neutrino sources.
We then use the latest measurement of the sub-PeV Galactic diffuse emission with HAWC
and \tas~to set competitive constraints on the ALP-photon coupling 
in a portion of the ALPs 
parameter space only mildly explored by gamma-ray telescopes so far.
To this end, we model the guaranteed diffuse gamma-ray contributions
from 
cosmic-ray interactions with gas and radiation, and from TeV emitters below the 
telescope's detection threshold.
This is the first time the model of Ref.~\cite{2017arXiv171201839V} is constrained with real data and that 
Galactic sub-PeV gamma rays are used to derive bounds on the ALPs' parameter space.

%The accompanying gamma-ray flux is naturally attenuated in the sources by interactions with internal radiation fields or -- if the sources are at high redshifts -- with the intergalactic medium. Photon-ALP conversion in the sources could convert a fraction of the gamma rays into ALPs which then travel unimpeded through the parent galaxy and reconvert in the Milky Way's magnetic field. 
%We aim at quantifying the cumulative ALP-induced gamma-ray flux from a population of extragalactic neutrino sources under certain assumptions about the physical environments of these objects as well as their redshift evolution 

The organization of the paper is the following one: 
In Sec.~\ref{sec:data}, we introduce the data set adopted and the characteristics more relevant for
our specific purpose. 
In Sec.~\ref{sec:astro_models}, we discuss the guaranteed contribution to the
sub-PeV gamma-ray diffuse emission coming from interstellar emission and unresolved sources.
The ALPs production model associated with high-energy neutrino emitting sources is described in Sec.~\ref{sec:alp_flux_model}.
After presenting the statistical framework for the analysis in Sec.~\ref{sec:statistics},
we then discuss our results in Sec.~\ref{sec:results} and conclude in Sec.~\ref{sec:discussion}.

%
%%%%%%%%%%%%%%%%%%%%%%%%%%%%%%%%%%%%%%
\section{The data}
\label{sec:data}
%%%%%%%%%%%%%%%%%%%%%%%%%%%%%%%%%%%%%%

The Tibet air shower and muon detector -- Tibet AS$\gamma$ in short -- has recently published the detection of sub-PeV gamma rays in the energy range from 100 TeV to 1 PeV originating in the Galactic disk \cite{2021PhRvL.126n1101A} and not associated with known, localized sources emitting in the TeV energy band. The collaboration reports their results for two regions of interest (ROI): \textit{(i)} $25^\circ < l < 100^\circ$ and $|b| < 5^\circ$ coinciding with the region in which the ARGO-YBJ collaboration reports the detection of a diffuse gamma-ray flux at TeV energies \cite{ARGO-YBJ:2015cpa}; \textit{(ii)} a larger region spanning $50^\circ < l < 200^\circ$ and $|b| < 5^\circ$, from which the CASA-MIA air shower array has derived upper limits on the gamma-ray emission towards the end of the past millennium \cite{Borione:1997fy}. 

The gamma rays seen by Tibet AS$\gamma$ are believed to be of Galactic origin, i.e.~to be almost exclusively due to hadronic cosmic-ray interactions within the Milky Way. 
Indeed, extragalactic contributions are highly suppressed at sub-PeV energies, because of the high opacity of the universe caused by the CMB and EBL \cite{Stecker:2016fsg, Franceschini:2017iwq}. The reported Tibet AS$\gamma$ diffuse emission measurement is also cleaned from events that are attributed to TeV-bright Galactic sources listed in the TeVCat online catalog\footnote{\url{http://tevcat2.uchicago.edu/}} and localized in the analysis ROI. This implies that Ref.~\cite{2021PhRvL.126n1101A} reports a true measurement of the diffuse flux along the Galactic plane:  The 38 gamma-ray events above 398 TeV are ``orphan'' events not associated with any known Galactic object.

We complement the Tibet AS$\gamma$ data set with the measurement of the Galactic diffuse emission reported by the HAWC collaboration~\cite{HAWC:2021bvb} 
in a region defined by $43^\circ < l < 73^\circ$ and $|b| < 2^\circ$ or $|b| < 4^\circ$ -- hence overlapping with the Tibet AS$\gamma$'s two ROIs -- and at energies from 10 TeV to 100 TeV, based on data taken between 2013 and 2019. Similarly to the Tibet AS$\gamma$ data set, the HAWC collaboration has derived the diffuse gamma-ray flux along the Galactic plane by subtracting the emission of all detected sources residing in the respective ROIs. 
The diffuse gamma-ray spectral energy distribution in both HAWC ROIs follows a power law with parameters quoted in Tab.~1 of Ref.~\cite{HAWC:2021bvb}.

In what follows, we will use both measurements of the truly diffuse emission from the Galactic plane (see Sec.~\ref{sec:statistics} for further details) in a combined analysis to constrain the coupling strength of ALPs to photons.

We notice that also the LHAASO-KM2A collaboration recently presented 
a preliminary measurement of the very-high-energy gamma-ray emission from the inner Galactic plane~\cite{Zhao:2021dqj}.
In this case, however, known Galactic sources are masked in the analysis.
Given the little information about the masking procedure and the very preliminary nature of the 
result, which for a few energy bins reveal some tension with \tas~data, we decided not to consider the LHAASO-KM2A measurement for the purpose of this work.
%[https://pos.sissa.it/395/859/pdf, https://icrc2021-venue.desy.de/video/Measurement-of-the-diffuse-gamma-ray-emission-from-Galactic-plane-with-LHAASO-KM2A/d52765cfda413f9fb395ae086afb80e4]

The adopted data are displayed in Fig.\ref{fig:spectral_plots}.

%%%%%%%%%%%%%%%%%%%%%%%%%%%%%%%%%%%%%%
\section{The astrophysical sub-PeV $\gamma$-ray diffuse flux} 
\label{sec:astro_models}
%%%%%%%%%%%%%%%%%%%%%%%%%%%%%%%%%%%%%%
To describe the conventional astrophysical gamma-ray emission along the Galactic plane at sub-PeV energies, we consider two ``guaranteed'' diffuse contributions: The interstellar emission (IE) and the cumulative emission from point-like and extended sources too faint to be resolved by the instruments whose measurements we are using, i.e.~sub-threshold (sTH).

\subsection{Interstellar emission model}
\label{sec:astro_iem}

The theoretical modeling of the IE at sub-PeV energies has recently gained increasing attention, 
and more and more refined models have consequently been published. 
This interest originates from a twofold reason: On the one hand, the availability of reliable measurements of the Galactic plane diffuse signal at these energies with current-generation instruments like HAWC \cite{HAWC:2021bvb}, Tibet AS$\gamma$ \cite{2021PhRvL.126n1101A} and LHAASO \cite{Zhao:2021dqj}, and, on the other hand, the need to push forward the theoretical understanding of this astrophysical component in order to compete with the growing experimental precision that is expected with the advent of LHAASO and Cherenkov Telescope Array (CTA).

Recently, Ref.~\cite{Luque:2022buq} has provided a comprehensive model for Galactic diffuse gamma-ray production at sub-PeV energies, investigating the effect of uncertainties related to cosmic-ray injection spectra and transport, and making use of latest available gas maps and cosmic-ray data.
Models with a spatial dependence of the diffusion coefficient 
rigidity index (``$\gamma$-optimized'' models) were found to better reproduce the observed hardening of cosmic-ray protons in \Fermi-LAT data, as well
as to match sub-PeV gamma-ray observations from ARGO-YBJ, HAWC, Tibet AS$\gamma$ and LHAASO better than conventional models 
where the diffusion coefficient rigidity behavior is constant in space. 
An alternative phenomenological model proposed in \cite{2018PhRvD..98d3003L}, and its comparison to Tibet AS$\gamma$ data, corroborates the claim of spatially dependent diffusion at sub-PeV energies.
In this analysis, we adopt two models (\texttt{MAX} and \texttt{MIN}) for the ``$\gamma$-optimized'' IE from \cite{Luque:2022buq}, which are a proxy of the uncertainty on the exact realization of the cosmic-ray transport in the Milky Way above 10 TeV while being consistent with cosmic-ray data. 
Absorption of gamma rays onto interstellar radiation fields and on the CMB is included. 

The gamma-ray maps, $\Phi^{\rm IE}(E, l, b)$, used in Ref.~\cite{Luque:2022buq} are publicly available.
We start from these data products to derive all results related to IE.
The IE \texttt{MAX} and \texttt{MIN} contribution in the two ROIs of interest for this
analysis are displayed in Fig.~\ref{fig:spectral_plots}. 

% Unresolved source contribution
\subsection{Contribution from unresolved sources}
The expected cumulative emission from a population of unresolved sources is a characteristic of each instrument as it depends on its specifications and performance. In other words, HAWC and Tibet AS$\gamma$ exhibit different detection thresholds  to gamma-ray sources within their sensitivity reach. 
Depending on the source-count distribution flux of the underlying population of gamma-ray sources, i.e.~the number of sources per unit flux $dN/dS$, sources too faint to be detected and whose flux 
is below the detection threshold will remain ``unresolved'' and contribute, cumulatively, to the diffuse gamma-ray emission.
The (true) underlying population of (Galactic) point-like and extended very-high-energy sources is then the same for both telescopes, so that a single phenomenological model to describe said population can be used.

%
%dN/dS discussion 
Studies of unresolved sources from different astrophysical objects
and their contribution to the gamma-ray diffuse emission 
have been extensively performed at GeV energies for \Fermi-LAT observations, e.g., \cite{Fermi-LAT:2015bhf},
especially to interpret the origin of the so-called \Fermi~diffuse gamma-ray 
background, see~\cite{Fornasa:2015qua} for a review. A quantitative characterization of the gamma-ray emission associated with unresolved sources in \Fermi-LAT data has been used to calibrate the IE models presented in Sec.~\ref{sec:astro_iem}.
While the contribution of 
unresolved sources is important for large-scale diffuse signals in extended
ROIs, as most of the \Fermi-LAT observations are, it is of lesser relevance
for typical Cherenkov telescopes measurements, given the limited 
instrument field of view (FOV).
Nonetheless, studies of bright TeV emitters have
shown that the corresponding unresolved population can significantly
contribute to the TeV large-scale diffuse signal in the Galactic disk~\cite{2008APh....29...63C,2019JCAP...12..050C, 2020A&A...643A.137S}.
The contribution of unresolved
sources is expected to rise at very high energies where 
large-FOV ground-based telescopes
are more sensitive to large-scale diffuse signals.

With the aim of quantifying what is the contribution 
of unresolved sources to the diffuse emission measurements, 
we follow the prescriptions outlined in \cite{2021arXiv210714584V}, whose details we further describe in Appendix \ref{app:subthresh_population}.

We consider that the Galactic population of very-high-energy sources is spatially distributed according to pulsars following the best-fit model of Ref.~\cite{Lorimer:2006qs}, with $\rho({\bf r})$ the source number density, c.f.~Eq.~\ref{eq:source_density}. 
We normalize $\rho\!\left(\bm{r}\right)$ to unity when integrated over the full volume of the Milky Way.

We model the gamma-ray luminosity function of TeV-bright sources, $\mathcal{L}(L_{\mathrm{TeV}})$, 
as in Ref.~\cite{Cataldo:2020qla}, where the function parameters are tuned to match the outcome of the H.E.S.S.~Galactic plane survey by assuming a population of pulsar wind nebulae, c.f.~Eq.~\ref{eq:luminosity_function}, and $L_{\mathrm{TeV}}$ refers to the gamma-ray luminosity of an individual object of the population in the energy band from 1 to 100 TeV.
The number of sources per unit volume and luminosity can then be written as:
\begin{equation}
\label{eq:source_density_main}
    \frac{\diff N}{\diff^3r\,\diff L_{\mathrm{TeV}}} = \rho\!\left(\bm{r}\right) \times \mathcal{L}(L_{\mathrm{TeV}})\mathrm{.}
\end{equation}

To compute the flux of unresolved sources in a given energy range from  
the source population luminosity function, one needs to assume an average source spectrum. 
As in Ref.~\cite{2021arXiv210714584V}, we parametrize the average source spectrum
by a power law with exponential cutoff:
\begin{equation}
\label{eq:avg_spectrum_main}
\varphi\!\left(E\right) = K_0 \left(\frac{E}{1\;\mathrm{TeV}}\right)^{-\beta}\exp{\left(-\frac{E}{E_{\mathrm{c}}}\right)}\mathrm{,}
\end{equation}
where the value of $K_0$ follows from the requirement that Eq.~\ref{eq:avg_spectrum_main} is normalized to one when integrated from 1 TeV to 100 TeV. The spectral parameters $\beta$ and $E_{\mathrm{c}}$ are free parameters of the model. 
We adapt the value of $\beta$ and $E_{\mathrm{c}}$ 
in order for our model to better match the properties of detected TeV-bright Galactic sources, as well as gamma-ray data from HAWC and Tibet AS$\gamma$, and therefore these parameters' values differ slightly from what has been employed in \cite{2021arXiv210714584V}.

The cumulative flux, $\Phi^{\mathrm{sTH}}$, of all TeV-bright sources below a detection threshold $S_{\mathrm{TH}}$ is defined as:
\begin{equation}
\label{eq:cumulative_flux_main}
\Phi^{\mathrm{sTH}}\!\left(E\right) = \varphi\!\left(E\right) \intop_{0}^{S_{\mathrm{TH}}} \Phi_{\mathrm{TeV}}\frac{\diff N}{\diff \Phi_{\mathrm{TeV}}}\,\diff\Phi_{\mathrm{TeV}}\mathrm{.}
\end{equation}
where $\diff N/\diff \Phi_{\mathrm{TeV}}$ is directly related to Eq.~\ref{eq:source_density_main} under the change of variable $L_{\mathrm{TeV}} = 4\pi d^2\Phi_{\mathrm{TeV}}\langle E\rangle$ and integrating out the spatial dependence. We refer the reader to Appendix~\ref{app:subthresh_population} for more details and definitions, which follow from Refs.~\cite{Cataldo:2020qla, 2021arXiv210714584V}.
The detection threshold, $S_{\mathrm{TH}}$, depends on the telescope performance. 
By comparing our model predictions from Eq.~\ref{eq:cumulative_flux_main} with HAWC and Tibet AS$\gamma$ published instrument performance,
we derive an estimate of each telescope's detection threshold $S_{\mathrm{TH}}$. 

We highlight now the details of our procedure to fix the values of the parameters
$S^{\rm Tibet}_{\mathrm{TH}}$, $S^{\rm HAWC}_{\mathrm{TH}}$, $E_c$ and $\beta$. 

\begin{figure}[t!]
\vspace{0.cm}
\includegraphics[width=0.975\columnwidth]{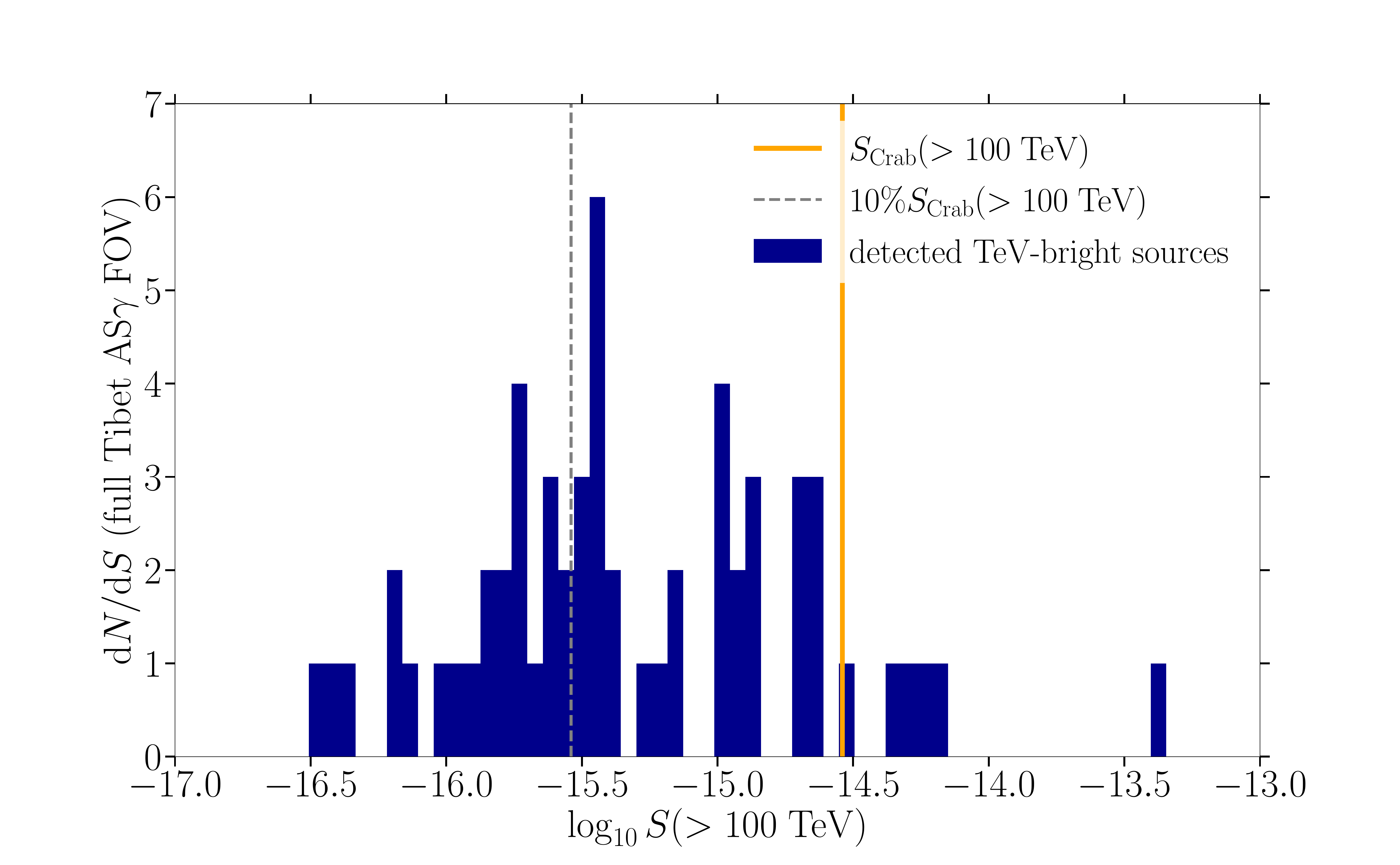}\\
\includegraphics[width=0.975\columnwidth]{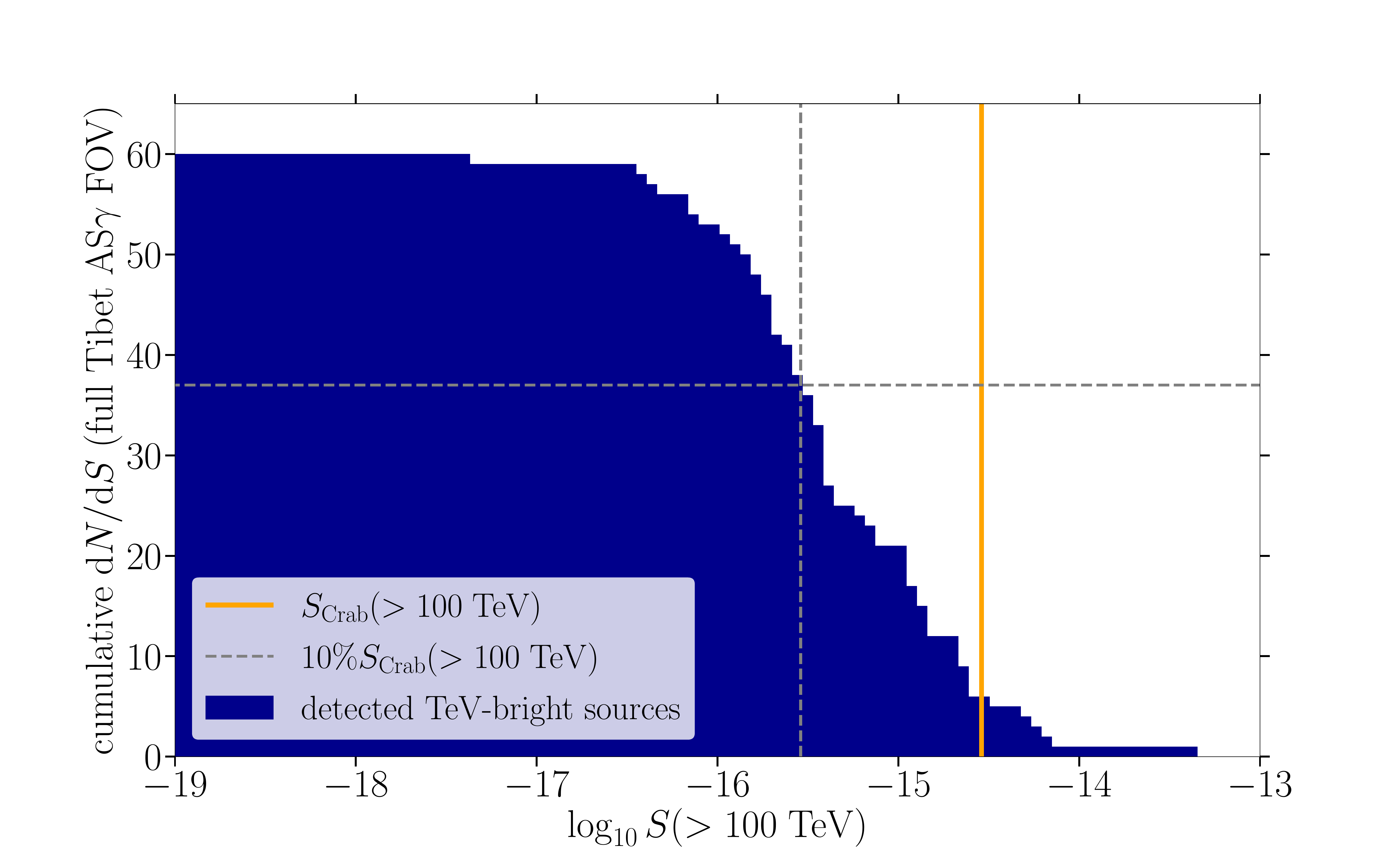}
\caption{(\emph{Top}:) Differential number of sources listed in TeVCat that fall within the FOV of Tibet AS$\gamma$ depending on their flux above 100 TeV $S(> 100\;\mathrm{TeV})$. (\emph{Bottom}:) Same as top panel but displaying the cumulative number of sources. The spectral information for each source has been extracted either directly from TeVCat or the associated references therein. For comparison, we display in orange $S(> 100\;\mathrm{TeV})$ of the Crab nebula according to its spectrum reported in Ref.~\cite{2019ApJ...881..134A}, as well as the fiducial threshold of Tibet AS$\gamma$ (vertical grey line) defined by requiring that at most 37 sources could have contributed one photon to the total excess counts obtained from stacking the observations in the direction of all known TeV-bright sources in the full FOV of the array. 
\label{fig:dNdS_tibet}}
\end{figure}

For Tibet AS$\gamma$, the collaboration reports that a dedicated search towards the known positions of 60 TeVCat sources within their total FOV ($22^{\circ} < \ell < 225^{\circ}$, $|b| < 5^{\circ}$) resulted in 37 gamma-ray-like events ($E > 100$ TeV) when applying a $0.5^{\circ}$ search window. \footnote{This number does not account for subtraction of background estimated events. 
Considering 8.7 background events, the source-like event counts reduce to 28.3. We notice that 
using this number instead of 37 would imply a larger detection threshold, i.e.~a larger contribution
of unresolved sources to the diffuse Galactic signal, and, ultimately, a stronger bound on the ALP-photon coupling.}
We interpret this stacked number of source-like excess counts in a conservative way:
We assume that at most 37 of these known sources could have contributed to the measured emission with exactly a single gamma ray. Hence, the Tibet AS$\gamma$ array has -- in the most optimistic scenario -- a sensitivity suited to detect the 37 brightest sources listed in TeVCat within its FOV. 
We analyze the properties of the currently detected TeV-bright Galactic sources in the TeVCat that fall within the total Tibet AS$\gamma$ FOV, and use the reasoning above to set a lower bound on the experimentally achievable detection threshold -- this is a conservative choice for the final purpose of this work as explained in detail below. 
From TeVCat and the listed references therein, we collect the spectral parameters of the known TeV-bright sources in the total FOV of Tibet AS$\gamma$.
%
%We note that the LHAASO collaboration recently announced detection of a number of Galactic sources\cite{2021Natur.594...33C}, but did not provide spectral characterization for the majority of them, only assessing what is the differential photon flux at 100 TeV in units of the Crab nebula's flux.
%
Among the 60 TeVCat sources, there are also a number of Galactic objects
whose detection has been recently announced by the LHAASO collaboration~\cite{2021Natur.594...33C}.
Unfortunately, most of these sources lack spectral characterization, and only 
the differential photon flux at 100 TeV in units of the Crab nebula's flux is provided.
We include these LHAASO sources assuming that they exhibit a differential gamma-ray spectrum coinciding with the one of the Crab as stated in Ref.~\cite{2019ApJ...881..134A}, and rescale the flux normalization according to the given photon flux information. 
Additionally, since the Tibet AS$\gamma$ collaboration does not find any gamma-ray event above 400 TeV associated with the positions of detected TeV-bright sources, we impose an exponential cutoff at $E_c = 300$ TeV for all TeVCat sources whose spectrum is described by a simple power law. We notice that the selected cutoff energy is in line with the reported spectra of the three PeVatron candidates detected by LHAASO \cite{2021Natur.594...33C}. 
We then calculate the flux $S$ above 100 TeV for each one of the 60 TeV-bright sources considered.
We show the cumulative source count distribution, $\mathrm{d}N/\mathrm{d}S$, of the 60 TeVCat sources in the bottom panel of Fig.~\ref{fig:dNdS_tibet}.
By imposing that at least 37 sources have been detected by \tas, we conclude that the detection threshold of Tibet AS$\gamma$ cannot be smaller than $10\%$ of the Crab nebula's flux above 100 TeV, which we adopt as the value for $S^{\rm Tibet}_{\mathrm{TH}}$. 

In case of HAWC, instead, the source detection threshold is obtained from the second HAWC catalog (2HWC)~\cite{Abeysekara:2017hyn} by deriving the flux in the energy range of interest, from 10 to 100 TeV, for each listed source and setting $S^{\rm HAWC}_{\mathrm{TH}}$ to the minimal value among the resulting flux values, i.e.~$2\%$ of the Crab flux in this energy range.

Lastly, we compute the average value of the spectral index for the set of detected TeV-bright sources in Tibet AS$\gamma$'s FOV -- which includes the HAWC ROI as a subset --, and use it as the average spectral index for the synthetic source population,  $\beta = 2.6$. 
We set the energy cutoff of the synthetic source population $E_c$ to 300 TeV, consistently with 
what assumed for the TeV-bright emitters.

We show the resulting cumulative flux of sub-threshold point-like and 
extended sources, $\Phi^{\rm sTH}$, in Fig.~\ref{fig:spectral_plots}, computed from Eq.~\ref{eq:cumulative_flux_main} for the \tas~and HAWC ROIs and derived instrument thresholds. We find that in the HAWC energy range $25\%$ of the total observed emission is due to resolved sources (taken from \cite{HAWC:2021bvb}) while IE accounts for $56\% / 50\%$ and sub-threshold sources for $44\% / 50\%$ in the \texttt{MAX}/\texttt{MIN} scenario. In the case of Tibet AS$\gamma$, we obtain that IE contributes $70\%/56\%$ and sub-threshold sources about $30\%/44\%$ (\texttt{MAX}/\texttt{MIN}) to the measured gamma-ray flux excluding resolved sources.

%%%%%%%%%%%%%%%%%%%%%%%%%%%%%%%%%%%%%%
\section{The ALPs sub-PeV $\gamma$-ray diffuse flux} 
\label{sec:alp_flux_model}
%%%%%%%%%%%%%%%%%%%%%%%%%%%%%%%%%%%%%%
An exotic large-scale diffuse signal can eventually contribute 
to the diffuse sub-PeV gamma-ray emission.
We here consider the cumulative gamma-ray flux from ALP-photon conversion as sourced
by high-energy neutrino emitters.
We follow the prescriptions presented in \cite{2015arXiv151101530K, 2017arXiv171201839V}, 
considering the latest determination of the astrophysical neutrino flux and state-of-the-art values
for all the relevant parameters at play, e.g.~star formation rate, magnetic field distributions, etc.

\subsection{Modeling the sub-PeV gamma-ray emission of star-forming galaxies}

Given the constraints from GeV data on the contribution of $p-p$ production to the 
neutrino flux, we assume that the entirety of the observed astrophysical neutrino flux is generated by $p-\gamma$ interactions in extragalactic sources. We thereby follow and adopt a proposition that has already been studied in Ref.~\cite{Murase:2015xka} and which was corroborated by further evidence in Ref.~\cite{Capanema:2020rjj}.
$p-\gamma$ interactions generate an associated {\it in situ} gamma-ray spectrum, whose profile is closely related to the neutrino spectrum via \cite{2008PhRvD..78c4013K}
    \begin{equation}
        E_{\nu}^2\frac{\mathrm{d}N_{\nu}}{\mathrm{d}E_{\nu}}\left(E_{\nu} = \frac{E_{\gamma}}{2}\right) = \frac{3}{2}E_{\gamma}^2\frac{\mathrm{d}N_{\gamma}}{\mathrm{d}E_{\gamma}}
    \end{equation}
based on the relation $E_{\gamma}\approx 2E_{\nu}$, and on the fact that, in each interaction, two photons are produced per three neutrinos.
Another crucial constraint on our model comes from the extragalactic gamma-ray background (EGB) measured by \Fermi-LAT \cite{Fermi-LAT:2014ryh}. The EGB comprises all gamma-ray emission from extragalactic sources including resolved localized emitters after subtracting the MW's foreground contribution. It is dominated by the cumulative emission of resolved extragalactic sources like blazars \cite{Fermi-LAT:2015otn, DiMauro:2017ing}. A related quantity is the isotropic diffuse gamma-ray background (IGRB), which entails all extragalactic gamma-ray emission but the resolved emitters. In contrast to the EGB, the bulk of its emission is likely to be caused by star-forming galaxies \cite{Roth:2021lvk}. Both quantities can already be explained with gamma-ray emission arising from $p-p$ interactions in the relevant extragalactic sources. Hence, we assume that the gamma-ray flux due to the \pg~contribution follows a smoothly broken power law -- and hence the neutrino component as well -- to suppress its intensity at GeV energies. Following~\cite{2017arXiv171201839V}, we remain agnostic about the nature 
of the neutrino emitters, rather we assume that the underlying {\it in situ} neutrino spectrum is parameterized as
\begin{equation}
    \label{eq:source_spectrum}
        \frac{\mathrm{d}N_{\nu}}{\mathrm{d}E_{\nu}} = N_0 \left[\left(\frac{E_{\nu}}{E_b}\right)^2 + \left(\frac{E_{\nu}}{E_b}\right)^{2\alpha} \right]^{-\frac{1}{2}}\rm{.}
\end{equation}
In the present work, we fix $\alpha = 2.87$ to the best-fit power-law spectral index derived from the measured 7.5-year IceCube neutrino flux of astrophysical origin based on the high-energy starting event sample (HESE) \cite{2021PhRvD.104b2002A}. The break energy $E_b$ follows from a fit of Eq.~\ref{eq:source_spectrum} to the 7.5-year HESE and 6-year Cascade data \cite{IceCube:2020acn}. We find $E_b = 25$ TeV to provide an adequate fit to both data sets.
The spectrum normalization $N_0$ (with respect to the extragalactic source) is fixed by requiring that the theoretically predicted cumulative differential neutrino flux at Earth \cite{2015arXiv151101530K}
    \begin{equation}
    \label{eq:cosmo_neutrino_flux}
    \frac{\mathrm{d}\Phi_{\nu}}{\mathrm{d}E_{\nu}} = \frac{c}{4\pi}\intop_0^{\infty} \frac{\mathrm{d}N_{\nu}}{\mathrm{d}E^{\prime}_{\nu}} (1 + z)\dot{\rho}_{\ast}(z)\left|\frac{\mathrm{d}t}{\mathrm{d}z}\right|\,\mathrm{d}z\rm{,}
    \end{equation}
yields the measured 7.5-year HESE neutrino flux (single flavor) of $2.12\times10^{-18}\;\mathrm{GeV}^{-1}\,\mathrm{cm}^{-2}\,\mathrm{s}^{-1}\,\mathrm{sr}^{-1}$ at 100 TeV \cite{2021PhRvD.104b2002A}. In Eq.~\ref{eq:cosmo_neutrino_flux}, $E^{\prime}_{\nu} = E_{\nu} (1+z)$, $\left|\mathrm{d}t/\mathrm{d}z\right| = H_0^{-1} (1 + z)^{-1} \left(\Omega_{\Lambda} + \Omega_m (1+z)^3\right)^{-1/2}$ with $H_0 = 70\;\mathrm{km}\,\mathrm{Mpc}^{-3}\,\mathrm{s}^{-1}$ and $\Omega_m = 0.3$ as well as $\Omega_{\Lambda} = 0.7$.
The functional parameterization of the star-formation rate density $\dot{\rho}_{\ast}(z)$ at redshift $z$ is taken from Ref.~\cite{Yuksel:2008cu}. We adopt the therein reported parameter values as benchmark scenario, while we employ the variations of these parameters proposed in Ref.~\cite{Horiuchi:2008jz} (for the Salpeter initial mass function, IMF) in order to study the impact of the uncertainty associated to this quantity. The choice of the cosmological parameters above is consistent with this prescription for $\dot{\rho}_{\ast}$. As a result, we obtain the following normalization constants (units of $10^{-19}\,\mathrm{GeV}^{-1}\,\mathrm{cm}^{-2}\,\mathrm{s}^{-1}\,\mathrm{sr}^{-1}$): (benchmark scenario) $N_0 = 2.42$; (upper boundary) $N_0 = 2.67$; (lower boundary) $N_0 = 2.13$.
We anticipate that uncertainties in the IMF will have a negligible impact on our final limits.

\subsection{From gamma rays to ALPs in star-forming galaxies }

Gamma rays escaping the neutrino sources, under the hypothesis 
of a non-zero ALP-photon
coupling, can then convert into ALPs and back in the presence of external 
magnetic fields.
We consider efficient ALP-photon conversion {\it within} extragalactic sources and in the Milky Way, whereas we neglect any potential conversion within the intergalactic medium \cite{Kartavtsev:2016doq} which follows from the assumption that the extragalactic magnetic field does not attain values around the observational upper bounds of $B\lesssim6\times10^{-9}$ G for a turbulent magnetic field with coherence length of 50 Mpc or $B\lesssim1\times10^{-9}$ G for a coherence length corresponding to the Hubble horizon based on Faraday rotation of light from distant quasistellar objects \cite{Blasi:1999hu}. The Planck collaboration has quantified the upper bounds on the extragalactic magnetic field from Faraday rotation of CMB photons amounting to $B\lesssim1380$ nG for a comoving coherence length of 1 Mpc \cite{Planck:2015zrl}.
The details of the ALP propagation formalism can be found in the literature, 
see, for instance, Refs.~\cite{DeAngelis:2011id, Kartavtsev:2016doq}. In Appendix~\ref{app:gamma_ALP_mixing}, we
summarize some basic equations related to this formalism, so that 
the reader can more easily follow our discussion and grasp the relevance of
the different astrophysical parameters at play. 
All calculations of the ALP-photon conversion probability either in extragalactic or Galactic environments are performed with the publicly available python package \texttt{gammaALPs} \cite{meyer_manuel_2021_6344566}, which provides the functionalities to apply the domain model approximation or the conversion in the Galactic magnetic field. In all the required calculations, we employ a user-defined photon-photon dispersion relation (based on the choice of interstellar radiation fields below) following the prescription in Ref.~\cite{Dobrynina:2014qba}.

The photon-ALP conversion probability {\it within} extragalactic sources depends on intrinsic properties like their magnetic field strength $\bm{B}$, their electron density and the strength of their interstellar radiation fields.
While, up to this point, no assumption on the neutrino source has been made -- since
we calibrated the gamma-ray flux at production using the {\it observed} neutrino flux --, 
a minimal degree of specification is required to compute the photon-ALP conversion probability
within these sources.
Among others, star-forming galaxies have been recently advocated to generate a sizable contribution to the observed diffuse neutrino flux~\cite{2020MNRAS.493.5880P, 2021MNRAS.503.4032A, 2022MNRAS.511.1336P}, despite early reported discrepancies between neutrinos produced by hadronic processes (in particular $p-p$ interactions) in star-forming galaxies and \Fermi-LAT constraints on their companion gamma-ray emission. \footnote{Taking into account the gamma-ray contribution from the cosmic-ray production mechanisms presented in some of these more recent works may further enhance the expected ALP-induced gamma-ray flux from star-forming galaxies with respect to what is presented here. A detailed and thorough analysis of the combined diffuse ALP flux from $p-p$ and \pg~interactions is left for future study.}

%
%Despite the reported discrepancies between neutrinos produced by hadronic processes (in particular $p-p$ interactions) in star-forming galaxies and \Fermi-LAT constraints on their companion gamma-ray emission, this source population has recently been the subject of revived interest. Depending on the invoked astrophysical environments and cosmic-ray interaction channels, star-forming galaxies have been advocated to generate a sizable contribution to the observed diffuse neutrino flux~\cite{2020MNRAS.493.5880P, 2021MNRAS.503.4032A, 2022MNRAS.511.1336P}, which consequently increases the level of the concomitant gamma-ray flux in the sub-PeV energy range compared to previous estimates. Taking into account the gamma-ray contribution from the cosmic-ray production mechanisms presented in these works, may hence further enhance the expected ALP-induced gamma-ray flux from star-forming galaxies. A detailed and thorough analysis of the combined diffuse ALP flux from $p-p$ and \pg~interactions is left for future study.
%
We here resort to a modeling based on a prototypical star-forming galaxy with parameters 
in accordance with observations of nearby galaxies, see Ref.~\cite{2019Galax...8....4B} for a review
and references therein.
%similar to the Milky Way at redshift $z = 0$ and a parametric function to describe the evolution with increasing redshift of the parameters at play. 
In particular, to model the dispersion effects, we assume an electron density of $n_e = 0.05\;\mathrm{cm}^{-3}$ 
%($10\%$ of the interstellar medium's particle density) 
at $z = 0$~\cite{2019Galax...8....4B}, and apply the redshift evolution in Ref.~\cite{2016ApJ...827..109S} to describe sources at higher redshifts. 
%\FC{[Assuming a constant filling factor, $\langle n_e \rangle \propto \langle B_{\rm tot} \rangle^{1.5}$~\cite{2019Galax...8....4B}. They found $\langle n_e \rangle \sim 0.05$, quite lower. From Tab.3 ]}
We adopt the selection of radiation fields proposed in Ref.~\cite{2015MNRAS.446....2S, 2016ApJ...827..109S} -- which includes an ultraviolet, optical and infrared radiation field as well as the CMB -- and the associated redshift evolution to model these quantities in our population of extragalactic sources.

A pivotal question in our model is the strength and structure of the magnetic field typically realized in the extragalactic neutrino sources under study. 
We are interested in modeling the large-scale, coherent or regular, magnetic field in the neutrino sources.
We rely on the recent results of the CHANG-ES survey of nearby disk and spiral galaxies~\cite{2012AJ....144...43I}.
Ref.~\cite{2020A&A...639A.112K} analyzes the magnetic fields in the halo of spiral galaxies.
The polarization stacking analysis of 28 galaxies shows a clear
underlying, ordered, X-shaped structure, which emerges as a common
feature of the mean magnetic field of likely most spiral galaxies. 
A regular, large-scale magnetic field in the halo is detected
in 16 galaxies, extending at least up to 7 kpc distance from the midplane of the disk. Nonetheless, the radio polarization measurements provide indications that the extension of the ordered magnetic fields is larger. This is
consistent with the typical assumption of energy equipartition between magnetic field and non-thermal electrons, which implies
that the magnetic field scale height
is a factor of two to four larger 
than the measured synchrotron halo height. 
Ref.~\cite{2020A&A...639A.112K} also concludes that the large-scale (coherent) magnetic fields in the halos are ordered on scales of about 1 kpc or larger.
We model the magnetic halo of the sources to be spherically symmetric and 
extending up to 10 kpc from their center. We fix the coherence length to $L_0 = 1$ kpc.
%(in accordance with the Milky Way prototype model proposed in Ref.~\cite{2016ApJ...827..109S} and a coherence length of $L_0 = 1$ kpc. 
The coherence length is relevant for the calculation of the photon-ALP conversion probability. To this end, we employ the domain model approximation \cite{DeAngelis:2007dqd, Mirizzi:2009aj, DeAngelis:2011id} in which the conversion probability is evaluated for a series of cells with fixed length and magnetic field conditions (strength and orientation with respect to the photon/ALP propagation direction). 
We use the mean regular magnetic field strength and associated variance derived from a sample of 21 star-forming galaxies~\cite{Fletcher:2011fn}, i.e.~$|\bm{B}| = 5 \pm 3$ \textmu G at $z=0$ to quantify the distribution of the magnetic field strength perpendicular to the line-of-sight $B_{\perp}$. 
In each cell of the domain model approximation, the components of the magnetic field vector $\bm{B}$ are drawn from a Gaussian distribution with zero mean and variance $2B_{\perp}^2/3$ to ensure that $\langle|\bm{B}_{\perp}|\rangle \approx B_{\perp}$.

The redshift evolution of magnetic field extension and strength in extragalactic sources suffers from uncertainties similar to the cases of magnetized jets or galaxy clusters. While coherent magnetic fields have been observed in distant galaxies, for instance at $z = 0.439$ as reported in Ref.~\cite{2017NatAs...1..621M}, their generation and evolution are still a subject of ongoing research from a numerical \cite{2019MNRAS.483.2424R, Pfrommer:2021uzq} as well as an observational \cite{2019Galax...7...54K} point of view. There are indications that large-scale regular galactic magnetic fields require time to form, rendering their existence a feature that is expected for galaxies at $z < 1$ \cite{Arshakian:2008cx} whereas small-scale turbulent magnetic fields should dominate at higher redshifts. Likewise, it is not fully understood if and how much the magnetic field strength changes with increasing redshift, an uncertainty also pointed out in our initial reference \cite{2016ApJ...827..109S}. To estimate the impact of the magnetic field modeling on our constraints, 
we consider two scenarios: \emph{(i)} The redshift evolution of the magnetic field strength is adopted from the prescription of Eq.~9 in Ref.~\cite{2016ApJ...827..109S}, while the coherence length $L_0$ scales as $L(z) = L_0 / (1+z)$; \emph{(ii)} The field strength of the regular magnetic field does not change compared to its values at $z=0$ but we keep the redshift evolution $L(z)$.

For both scenarios, we derive the average expected ALP flux from the extragalactic neutrino source population at redshift $z$ by incorporating the uncertainty on $B_{\perp}$. We determine the sample averages over 5000 single realizations of extragalactic sources where $B_{\perp}$ is drawn from a Gaussian distribution with mean and variance corresponding to the experimentally inferred values above.

Besides the large-scale regular field, most disk and spiral galaxies
also show evidence for turbulent magnetic fields at few hundred parsecs scales,
with intensities compatible with the regular ones.
%
%Even more so, galaxies at high redshifts are likely to undergo a phase of starburst activity in their core region \cite{1982A&A...105..342L}, which induces the creation of strong, turbulent magnetic fields \cite{Sudoh:2018ana, 2019MNRAS.487..168P, Pfrommer:2021uzq}. However, the turbulent features reduce with increasing distance from the starburst region, as pointed out via the example of M82 \cite{2017A&A...608A..29A} or magneto-hydrodynamical simulations \cite{Pfrommer:2021uzq}. 
%In its halo, the regular component of the magnetic field seems to follow the basic assumptions we have made for our source population. 
%
We checked that the contribution of the turbulent component is 
negligible in our model, as also pointed out in Ref.~\cite{Carenza:2021alz} for the Milky Way magnetic field.
We therefore neglect the turbulent magnetic field component.

\begin{figure}[t!]
\vspace{0.cm}
\includegraphics[width=0.975\columnwidth]{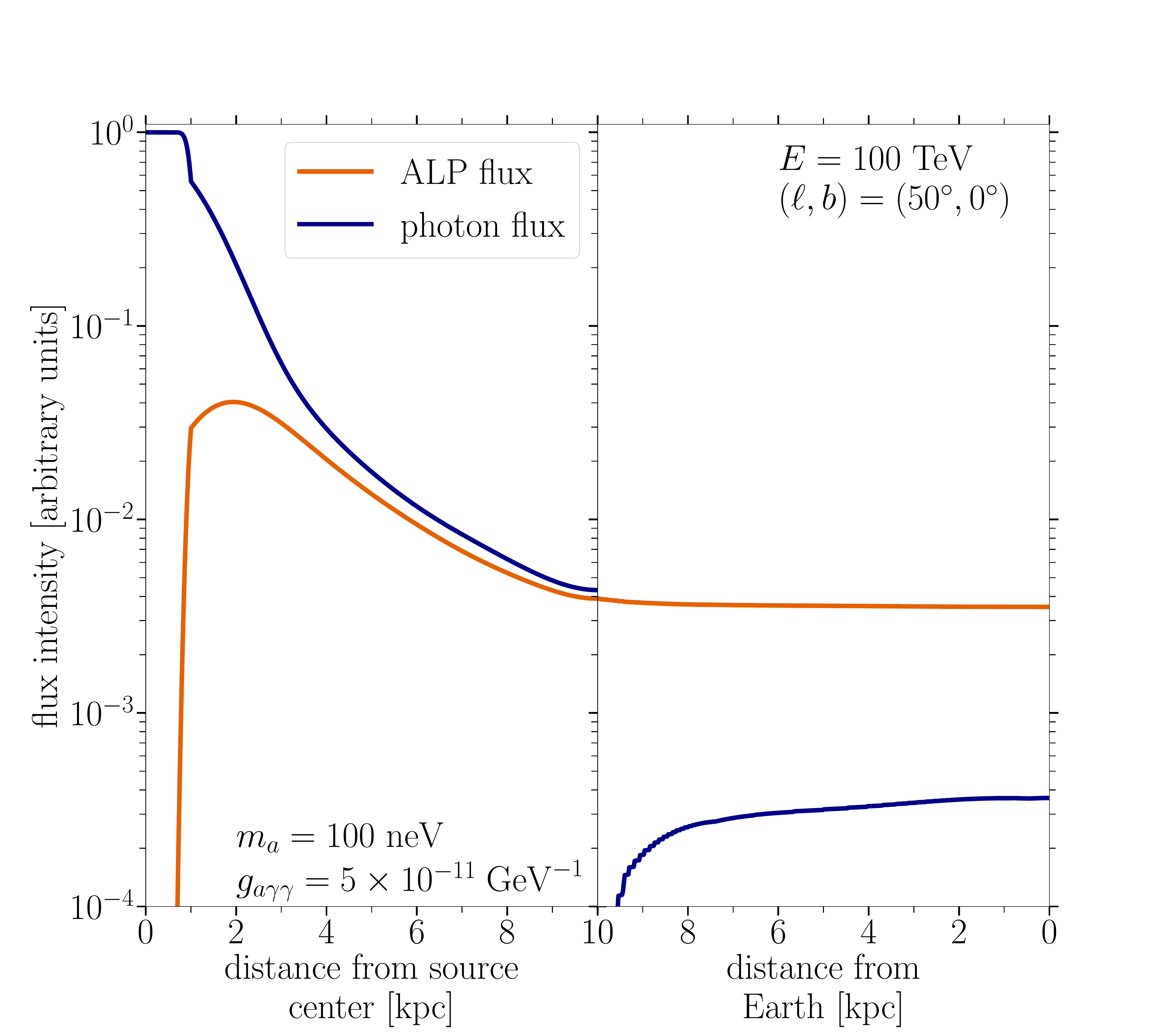}
\caption{Evolution of the photon (blue) and ALP (orange) fluxes as a function of the distance traveled within either an extragalactic source (left panel) or the Milky Way (right panel). The displayed photon flux has been normalized to one, the ALP flux scales accordingly. For definiteness, the ALP parameters are set to $m_a = 100$ neV and $g_{a\gamma\gamma} = 5\times10^{-11}$ GeV$^{-1}$ while the considered energy of the particles is 100 TeV. To calculate the conversion within extragalactic objects, we have applied the approach for magnetic field scenario \emph{(i)} described in the text to derive an average expectation. The sources are located at redshift $z=1$. For the Milky Way, we have chosen the representative direction $\left(\ell, b\right) = \left(50^{\circ}, 0^{\circ}\right)$, which is part of the ROI of both instruments whose measurements we utilize to set upper limits on the ALP parameter space. \label{fig:flux_evolution}}
\end{figure}

\subsection{ALP re-conversion in the Milky Way and its implications}

A fraction of the ALP flux arriving at the Milky Way will reconvert into gamma rays due to the Galactic magnetic field. The structure and strength of its regular component -- at least in comparison to extragalactic magnetic fields -- are more precisely known. We utilize the Jansson \& Farrar Galactic magnetic field model \cite{2012ApJ...757...14J} with parameter updates according to the measurements of the Planck satellite \cite{2016A&A...596A.103P}. Such more detailed description of the magnetic field structure allows us to derive the probability for a reconversion event for each line-of-sight. Besides, the Milky Way hosts a number of interstellar radiation fields whose spatial distribution we describe with the profile reported in Ref.~\cite{Misiriotis:2006qq},  contributing to both dispersion and absorption terms, c.f.~Appendix~\ref{app:gamma_ALP_mixing}.
Their temperatures and relative normalizations are again taken from Ref.~\cite{2016ApJ...827..109S} to ensure consistency with the prototype model applied to the extragalactic source population.

As an illustrative example, we display in Fig.~\ref{fig:flux_evolution} the evolution of the photon (normalized to 1) and ALP fluxes ($m_a = 100$ neV and $g_{a\gamma\gamma} = 5\times10^{-11}$ GeV$^{-1}$) within extragalactic sources (left panel) and in the Milky Way (right panel) at a representative energy of 100 TeV. The extragalactic source panel represents the average result from an ensemble of individual objects at redshift $z=1$ where the photon-ALP conversion has been derived as outlined above. Concerning the Milky Way, we show the impact of photon-ALP conversion events along the line-of-sight in the direction of $\ell = 50^{\circ}$ and $b=0^{\circ}$, i.e.~a direction that is located within the ROI of both telescopes under study. We do not show the evolution of the fluxes from the sources at $z=1$ to the Milky Way since one of our main assumptions is that no efficient conversion takes place in this intermediate domain. Moreover, the {\it in situ} gamma-ray flux is expected to be completely attenuated either in the source or on the CMB and EBL. The left panel of this figure shows that at sub-PeV energies the absorption of gamma rays on especially the cosmic microwave background is quite important in the high-redshift universe so that our assumption appears reasonable.

In Fig.~\ref{fig:spectral_plots}, we show the spectral energy distribution of the 
derived ALP-induced gamma-ray flux from star-forming galaxies for an ALP mass of 100 neV 
and a coupling corresponding to the upper limit for the \texttt{MAX} IE model (see below). 

In case of a non-zero coupling between photons and ALPs, gamma rays produced in the Milky Way will also convert into ALPs and escape detection by gamma-ray telescopes, thus reducing the signal we want to constrain. This affects astrophysically produced gamma rays from collisions of very-high-energy cosmic rays with the interstellar medium or radiation fields, as well as gamma rays produced by localized point-like or extended objects such as supernova remnants, pulsars or pulsar wind nebulae. 
In order to include this effect and lacking a 3D model for the production of photons along the line of sight, we incorporate the photon survival probability due to ALPs in an effective way: We modify the standard gamma-ray spectrum of these contributions by multiplying the energy and position-dependent Galactic photon survival probability calculated with \texttt{gammaALPs} with the astrophysical model prediction derived under the assumption of $g_{a\gamma\gamma} \equiv 0$. 
In particular, for the IE contribution, we multiply the all-sky map of the IE model by the survival probability map,
\begin{equation}
    \label{eq:survival_prob_IE}
    \Phi^{\rm IE}_{\rm abs}(E, l, b) = \Phi^{\rm IE}(E, l, b) \times (1 - P_{a\gamma}(E, l, b)) \, ,
\end{equation}
and then derive the expected flux in the ROI. In this way, the spatial dependence of the probability is fully accounted for. 
For the sub-threshold contribution, instead, we calculate the on-average expected survival probability in the corresponding ROI and then multiply the sub-threshold contribution flux by this factor: \footnote{We notice that the survival probability does not strongly vary within the ROIs considered.}
\begin{equation}
    \label{eq:survival_prob_sTH}
    \Phi^{\rm sTH}_{\rm abs}(E) = \Phi^{\rm sTH}(E) \times \langle 1 - P_{a\gamma}(E, l, b) \rangle_{\rm ROI} \, .    
\end{equation}
The numerical results that we obtain from \texttt{gammaALPs} are derived by considering propagation through the entire Milky Way along the line-of-sight, while Galactic gamma rays may be produced all along this trajectory. 
Albeit this approach potentially overestimates the modulation of the astrophysically expected spectra rendering our constraints rather conservative, we anticipate that the error induced is negligible with
respect to other sources of uncertainty.
As an example, for the particular realization of ALP parameters in Fig.~\ref{fig:spectral_plots}, the average loss of gamma rays is only around 4\% in the ROI of Tibet AS$\gamma$ or HAWC, over the relevant energy band from 10 TeV to 1 PeV.

%\sout{In Fig.~\ref{fig:spectral_plots}, we show the spectral energy distribution of the 
%derived ALP-induced gamma-ray flux from star-forming galaxies for an ALP mass of 100 neV 
%and a coupling corresponding to the upper limit for the \texttt{MAX} IE model (see below). }
%\sout{To be more explicit on the level of conservatism regarding the effective treatment of ALP-photon conversions for the astrophysical components, we find, for the particular realization of ALP parameters in Fig.~\ref{fig:spectral_plots}, that the average photon survival probability in the ROI of Tibet AS$\gamma$ is not less than 96\% over the relevant energy band from 10 TeV to 1 PeV, i.e.~the average loss of gamma rays is only around 4\%.}

%%%%%%%%%%%%%%%%%%%%%%%%%%%%%%%%%%%%%%%%%
\section{Statistical framework} 
\label{sec:statistics}
%%%%%%%%%%%%%%%%%%%%%%%%%%%%%%%%%%%%%%%%%

We conduct a combined maximum likelihood analysis to constrain the diffuse ALP flux generated by the very-high-energy gamma-ray emission from \pg-interactions in the population of neutrino-generating extragalactic sources. We choose a $\chi^2$-function as the fundamental quantity to construct the likelihood function $\mathcal{L}$, which reads
\begin{equation}
\mathcal{L} = \prod_{j \in \{\mathrm{Tibet}, \mathrm{HAWC}\}}e^{-\chi_j^2}\rm{.}
\end{equation}
The $\chi^2$-function is a function of a single parameter, namely the normalization $\theta$ of the ALP contribution $\Phi^{\mathrm{ALP}}$ to the astrophysically expected gamma-ray emission in the Tibet AS$\gamma$ and HAWC ROI. Quantitatively, it is defined as
\begin{equation}
\label{eq:chi_square}
\chi_j^2\!\left(\theta\right) = \sum_k \frac{\left(\Phi^{\mathrm{ALP}}_{k}\!\left(\theta\right) + \Phi^{\mathrm{IE}}_{k}\!\left(\theta\right) + \Phi^{\mathrm{sTH}}_{k}\!\left(\theta\right) - \Phi_{j,k}\right)^2}{\sigma_{j,k}^2}\rm{,}
\end{equation}
where the index $k$ runs over the energy bins of each experimental data set $\Phi_j$, $\Phi^{\mathrm{IE}}$ denotes the gamma-ray emission associated to the interstellar emission as predicted by the two models that we adopt; $\Phi^{\mathrm{sTH}}$ is the gamma-ray component due to unresolved point-like and extended sources in the ROI of the respective instrument according to our reasoning detailed in the previous section and $\sigma_j^2$ refers to the variance of the respective experimental data for which we use the upper error margin in case of asymmetric error bars. As mentioned in Sec.~\ref{sec:data}, the HAWC data is provided in terms of a continuous power law. Thus, we bin the spectrum in five logarithmically spaced energy bins between 10 TeV and 100 TeV to apply the aforementioned formalism. We explicitly introduced the dependence on the ALP-photon coupling for all of our model components to emphasize the impact of conversion events even on astrophysically produced gamma rays. We set upper limits on the normalization of the ALP component using a log-likelihood ratio test statistic, which in this particular case reduces to the difference between $\chi^2$ functions according to
\begin{equation}
\Delta\chi^2 = \chi^2\!\left(\theta\right) - \chi^2(\hat\theta)\rm{,}
\end{equation}
where $\hat\theta$ denotes the best-fit value of the ALP flux normalization parameter minimizing the value of the $\chi^2$-function in Eq.~\ref{eq:chi_square}. Since $\Delta\chi^2$ is a function of a single degree of freedom, we find the upper limit on $\theta$ at a $95\%$ confidence level (C.L.) when it attains a value of 3.84 \cite{10.1093/ptep/ptaa104}\footnote{\href{https://pdg.lbl.gov/2020/reviews/rpp2020-rev-statistics.pdf}{PDG Review Statistics }, Table 40.2.}. The constraint on $\theta$ can directly be translated to an upper limit on the coupling strength between ALPs and photons $g_{a\gamma\gamma}$ by using a grid of representative coupling strength values for fixed ALP mass $m_a$, which we interpolate.

\medskip
%%%%%%%%%%%%%%%%%%%%%%%%%%%%%%%%%%%%%%
\section{Results} 
\label{sec:results}
\begin{figure*}[t!]
\vspace{0.cm}
\includegraphics[width=0.975\columnwidth]{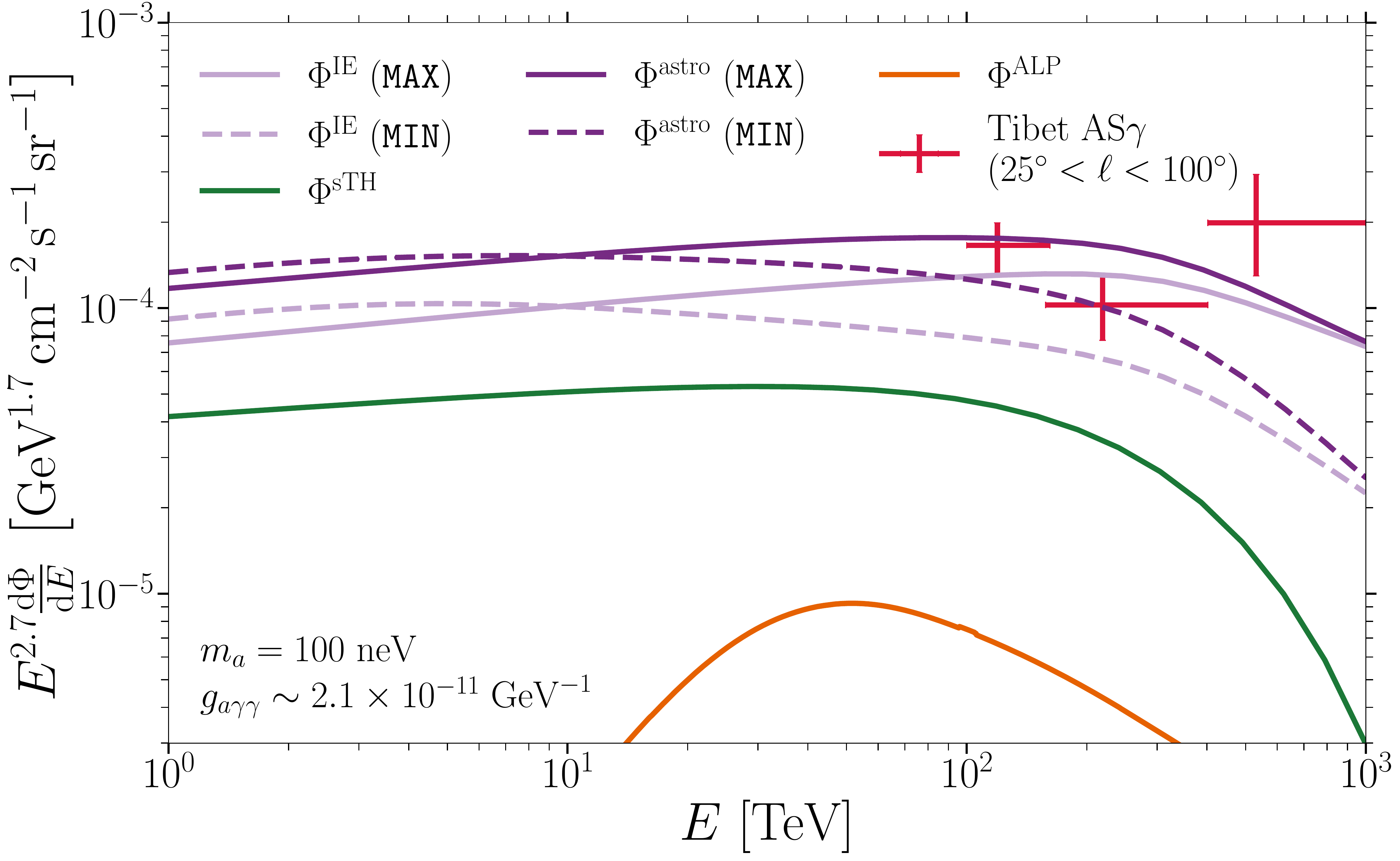}
\hfill
\includegraphics[width=0.975\columnwidth]{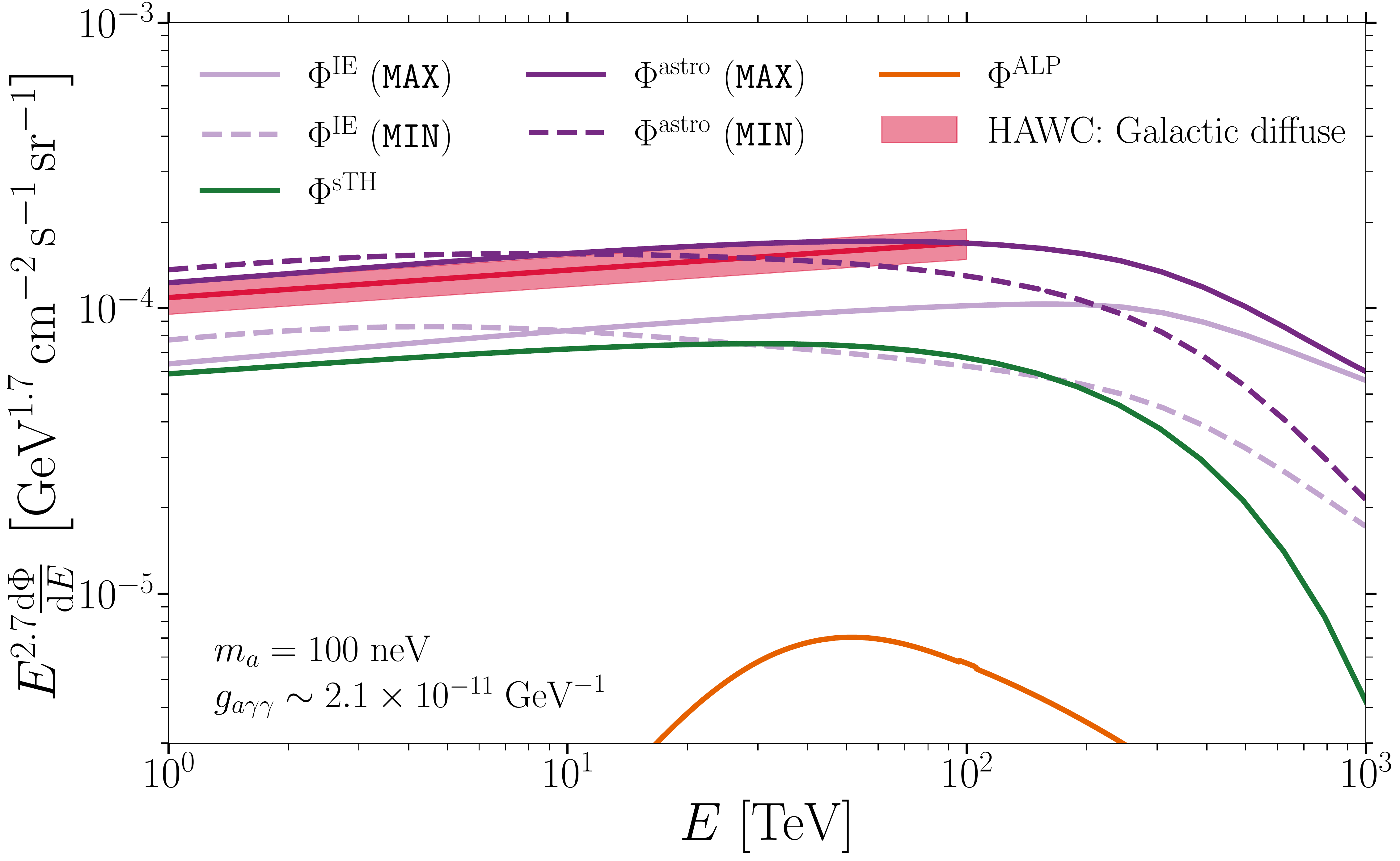}
\caption{Gamma-ray spectra of the emission components used to fit the Tibet AS$\gamma$ \cite{2021PhRvL.126n1101A} (ROI: $25^{\circ} < \ell < 100^{\circ}$, $|b| < 5^{\circ}$; \textit{left}) and HAWC data \cite{HAWC:2021bvb} (ROI: $43^{\circ} < \ell < 73^{\circ}$, $|b| < 4^{\circ}$; \textit{right}) of the Galactic diffuse emission (red). Light purple lines display the expected contribution of the interstellar emission (IE) either in a minimal (dashed) or maximal (solid) scenario whereas the component arising due to subthreshold sources is marked with a solid green line ($\alpha = -2.6$, $E_c = 300$ TeV, $S_{\mathrm{TH}}^{\mathrm{Tibet}} = 10\%S_{\mathrm{Crab}}(>100\;\mathrm{TeV})$, $S_{\mathrm{TH}}^{\mathrm{HAWC}} = 2\%S_{\mathrm{Crab}}(\left[10,100\right]\;\mathrm{TeV})$). The corresponding total astrophysical gamma-ray emission is denoted by dark purple lines adhering to the same IE line style. Note that we display here the theoretically predicted spectra for $g_{a\gamma\gamma}\equiv0$. For comparison, we add as an orange solid line an example of the derived ALP-induced gamma-ray flux normalized to the value corresponding to the upper limit in the \texttt{MAX} scenario for an ALP of $m_a = 100$ neV. 
\label{fig:spectral_plots}}
\end{figure*}

%%%%%%%%%%%%%%%%%%%%%%%%%%%%%%%%%%%%%%
The combined data from Tibet AS$\gamma$ and HAWC allow us to exploit the energy range from 10 TeV to 1 PeV to derive constraints on the parameter space of ALPs. After having conducted a maximum likelihood analysis, we find that the smaller ROI of Tibet AS$\gamma$ ($25^{\circ} < \ell < 100^{\circ}$, $|b| < 5^{\circ}$) combined with the larger ROI of HAWC ($43^{\circ} < \ell < 73^{\circ}$, $|b| < 4^{\circ}$) results in the most stringent upper limits on the ALP-photon coupling constant $g_{a\gamma\gamma}$ for all probed ALP masses.
In fact, when we only consider the theoretically modeled astrophysical contribution in both ROIs, as shown in Fig.~\ref{fig:spectral_plots} without any ALP-induced spectral modulation, the data is entirely consistent with having solely IE and an additional diffuse contribution from localized sources below the detection threshold of the respective instrument. 
As a useful measure to gauge the room left for an ALP signal (for ALP masses $m_a\lesssim 2\times10^{-7}$ eV) over the energy range of interest, 
we quote in Tab.~\ref{tab:ALPflux_vs_E} the maximally allowed ALP flux as a function of energy (adhering to the binning scheme employed to the HAWC flux and as stated by the Tibet AS$\gamma$ collaboration), for the different 
astrophysical background models adopted in this work. This information 
can consequently be used to recast our results to different models for the gamma-ray signal from ALP-photon conversion.

\begin{figure*}[t!]
\vspace{0.cm}
\includegraphics[width=0.975\columnwidth]{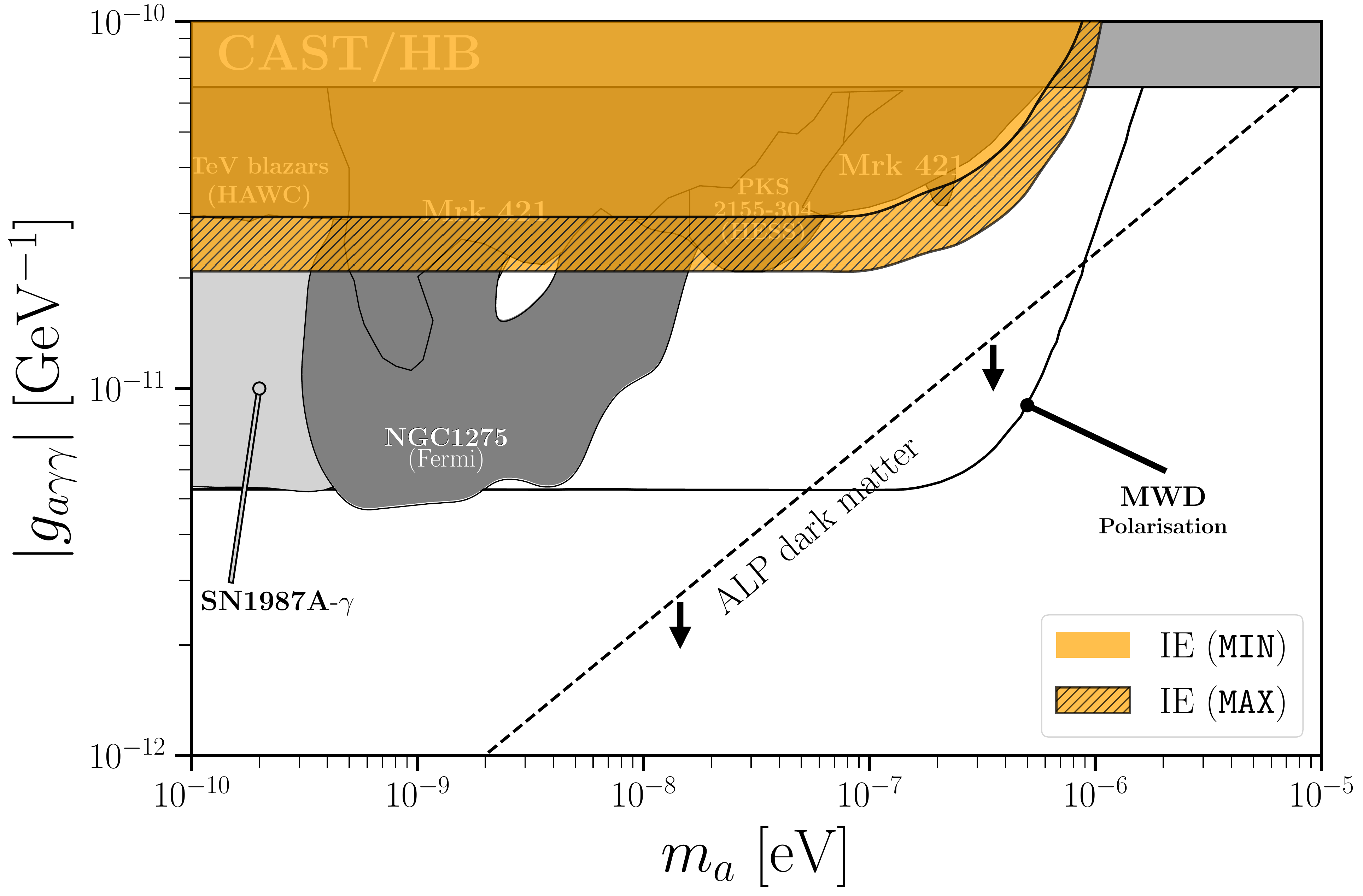}
\hfill
\includegraphics[width=0.975\columnwidth]{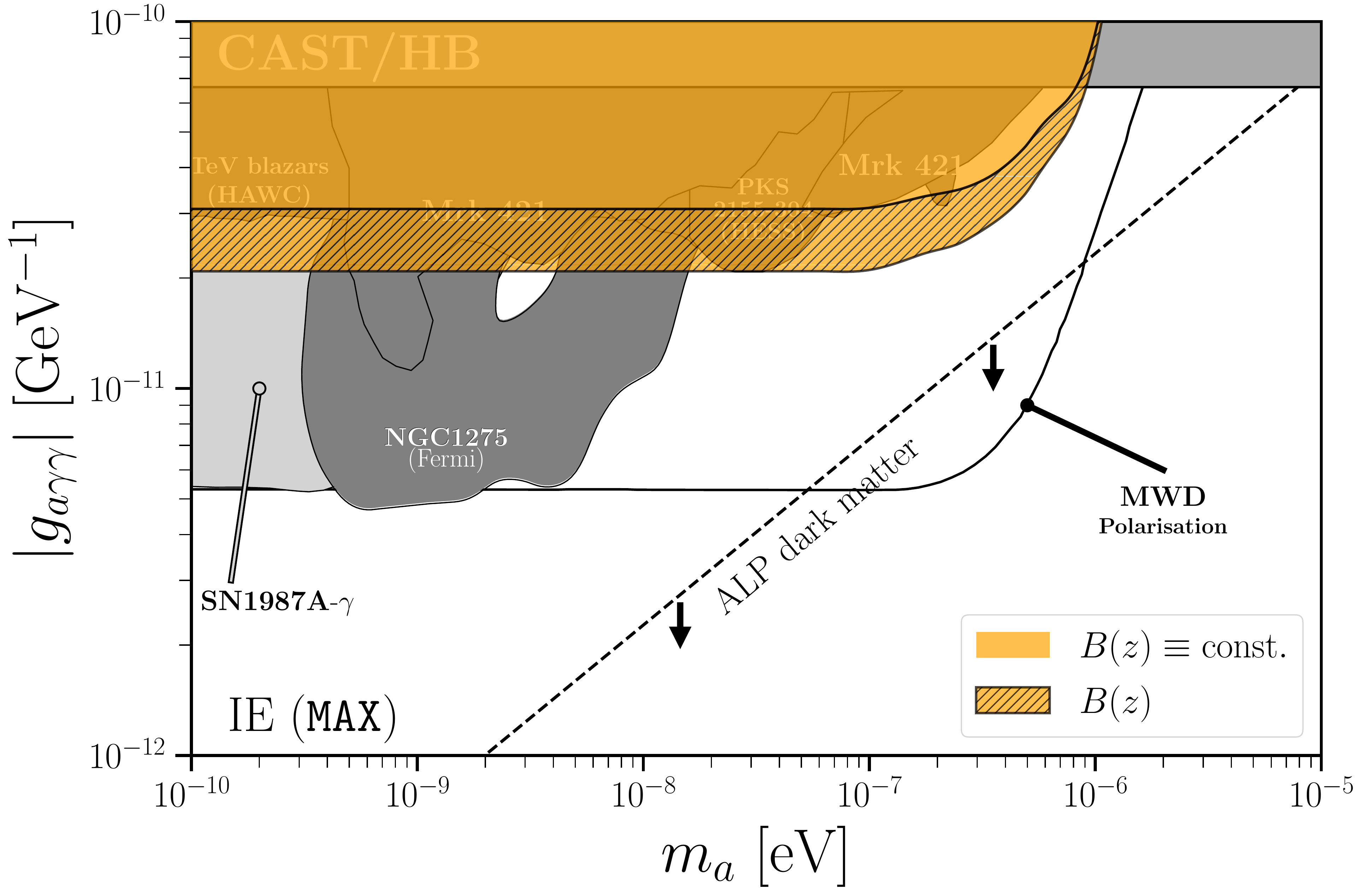}
\caption{$95\%$ C.L. upper limits on the ALP-photon coupling strength $g_{a\gamma\gamma}$ as a function of the ALP mass $m_a$ derived from the combined analysis of the Tibet AS$\gamma$ ($25^{\circ} < \ell < 100^{\circ}$, $|b| < 5^{\circ}$) and HAWC measurement ($43^{\circ} < \ell < 73^{\circ}$, $|b| < 4^{\circ}$) of the diffuse gamma-ray flux along the Galactic plane. \emph{Left:} Dependence of the upper limits on the choice of the IE model. The yellow-shaded region illustrates the constraints derived for the \texttt{MIN} scenario of the ``$\gamma$-optimized'' IE model from \cite{Luque:2022buq} while the enlarged black-hatched region denotes the improvement of upper limits if the \texttt{MAX} scenario of the same theoretical model is realized in nature. \emph{Right:} Uncertainty on the constraints arising from the two scenarios for the evolution of the magnetic field strength in extragalactic neutrino sources described in Sec.~\ref{sec:alp_flux_model}. The black-hatched  region displays the limit in scenario \emph{(i)} for the case of maximal IE (as in the left panel) whereas the yellow region illustrates the loss of sensitivity when instead scenario \emph{(ii)}, a constant magnetic field strength throughout the history of the universe, is assumed. For comparison, we show various constraints on ALPs derived from different observables relevant in the very-high-energy regime: HAWC TeV blazars \cite{Jacobsen:2022swa}, \Fermi-LAT measurements of the spectra of NGC 1275\cite{Fermi-LAT:2016nkz}, H.E.S.S.~searches for irregularities in the spectra of PKS 2155-304 \cite{HESS:2013udx}, combined ARGO-YBJ and \Fermi-LAT observations of Mrk 421 \cite{Li:2020pcn} as well as the non-observation of gamma rays from SN1987A \cite{Payez:2014xsa}. Besides these gamma-ray probes of ALP presence, we display the upper limits derived from the helioscope experiment CAST \cite{CAST:2017uph}, whose constraints overlap with an independent constraint from an analysis of the number of stars in the horizontal branch in old stellar systems \cite{PhysRevLett.113.191302} and a constraint due to the non-observation of polarization features in the emission of white dwarfs \cite{Dessert:2022yqq}. This plot and the collection of current ALP upper limits have been generated with the software and library provided by Ciaran O'Hare \cite{AxionLimits}.
\label{fig:ULIM_ALPs}}
\end{figure*}

We obtain competitive 95$\%$ C.L.~upper limits on $g_{a\gamma\gamma}$ as illustrated in Fig.~\ref{fig:ULIM_ALPs}. In the left panel, 
we show the variation of the limits induced by the change of the IE model, and
we confront our constraints with a sample of upper bounds derived from high-energy and very-high-energy gamma-ray instruments. We are able to improve some of these literature constraints for ALP masses $m_a > 10^{-8}$ eV for the maximal IE scenario.
We stress that the contribution from unresolved sources, at least for \tas, represents a lower limit of the unresolved source flux, because of the optimistic definition of the detection threshold. This is a conservative choice for our purposes, since
it leaves more space for ALPs and implies a weaker limit on the ALP-photon coupling.  

Quantitatively, we obtain in the case of the \texttt{MAX} IE model an upper limit of
\begin{equation}
\label{eq:upper_limit}
    g_{a\gamma\gamma} \lesssim 2.1\times10^{-11}\;\mathrm{GeV}^{-1} \,\,  \text{for} \,\, m_a \leq 2\times10^{-7} \, \rm eV \, .
\end{equation}
In the case of the \texttt{MIN} IE model, instead, the bounds degrade by a factor of $\sim 1.5$. We assume the magnetic field redshift evolution case \emph{(i)}.

In the right panel, we display the uncertainty in the limits due to the 
redshift evolution of the magnetic field in the neutrino sources, i.e.~
the non-trivial redshift evolution scenario ({\it i}) versus 
the constant 
magnetic field case ({\it ii}). In this latter case, 
the upper limit stated in Eq.~\ref{eq:upper_limit} degrades by about 50\%. 
This model ingredient is therefore a source of systematic uncertainty as relevant as the uncertainty of the IE at sub-PeV energies.

As anticipated, the uncertainty caused by the current imperfect knowledge of the star-formation rate density evolution $\dot{\rho}_{\ast}(z)$ is almost negligible, and accounts for a fractional change of $\sim\!2\%$ of the upper limits compared to the benchmark scenario. 
%\sout{exemplified via the upper limits for the minimal scenario of the chosen IE. The systematic uncertainty induced by this ingredient of our model is significantly lower \CE{-- in fact, almost negligible --} than the uncertainty due to the modeling of the IE, and accounts at most for a fractional change of $\sim\!2\%$ of the upper limits compared to the benchmark scenario.}

In the present analysis, we combine \tas~and HAWC data. By considering only one ROI at a time we, however, find that most of the constraining power is derived from the HAWC measurement, while the \tas~data set provides a less influential contribution.
%\FC{Instead of saying how much the limits from \tas~only are weak, we can positively rephrase the sentence below:}
Indeed, when including \tas~data we obtain an improvement of about a factor of 1.3.
The main reason is the strong \emph{in situ} absorption of gamma rays on CMB photons in the high-redshift universe depleting the expected ALP-induced gamma-ray flux at Earth at energies above 100 TeV. 
We notice, however, that in the future new measurements of the Galactic diffuse emission
at 10 TeV will turn to be truly complementary to measurements at lower energies. Consequently, these future sub-PeV data sets may even dominate the constraining power given that the peak of the ALP signal is indeed located at these energies.

%\sout{Standalone, \tas~yields a bound about about a factor of three weaker than the upper limit quoted in Eq.~\ref{eq:upper_limit}. The main reason for this observation is the strong \emph{in situ} absorption of gamma rays on CMB photons in the high-redshift universe depleting the expected ALP-induced gamma-ray flux at Earth at energies above 100 TeV. }
We note that -- despite the asymmetric importance of both data sets -- our results are not strongly sensitive to the assumed value of the break energy $E_b$ at tens of TeV used to fit the measured neutrino flux at Earth. To stress it once more, the break in the neutrino/gamma-ray spectrum reduces the impact of residual gamma rays that are -- despite our fundamental assumption -- not fully attenuated within the sources or on the CMB and EBL.

Finally, not accounting for photon losses regarding the astrophysical gamma-ray emission induced by conversion into ALPs would improve the limits by about 10\%. 

%
%Fig. 2: constraints gag - mass\\

\medskip
%%%%%%%%%%%%%%%%%%%%%%%%%%%%%%%%%%%%%%
\section{Discussion and conclusions}
\label{sec:discussion}
%%%%%%%%%%%%%%%%%%%%%%%%%%%%%%%%%%%%%%
We have presented the first analysis using sub-PeV (1 TeV -- 1 PeV) Galactic gamma rays to constrain ALPs coupling
to photons. 
We leveraged on the gamma-ray/neutrino multi-messenger connection and predict the cumulative gamma-ray flux of a population of high-energy astrophysical objects, which would be responsible for the entirety of the neutrino signal, and we include the effect of ALP-photon conversion in the sources, as well as in the Galaxy.
In order to model the guaranteed gamma-ray contribution to the Galactic diffuse emission
from ``standard'' astrophysics, we made use of the latest models for IE, which represent the state-of-art in the field, and carried out a careful analysis of sub-threshold point-like and extended sources by estimating realistic detection thresholds.
For this analysis, we combined \tas~and HAWC data.
In principle, also LHAASO preliminary data~\cite{Zhao:2021dqj} could have been added to the analysis further improving the limits.
Nonetheless, given the preliminary nature of these results, we preferred not to include them in this work.

In the mass range $m_a \sim 10^{-8} - 10^{-6}$ eV, 
the most constraining limits come from searches for ALP-induced polarization features in magnetic white dwarfs~\cite{Dessert:2022yqq}, with 
little astrophysical contamination to the linear polarization signal predicted so far.
Besides those, our limits result to be competitive with, and even supersede, most of the
current bounds from high-energy gamma-ray astrophysics.
%
%Besides those,  2. HAWC blazars depend on blazar jet magnetic field, intra-cluster magnetic field -- known to weaken the limits by o.d.m for NGC1275 --, and the MW magnetic field 3. Mrk 421 spectral features: B field in the jet, especially radius at which the jet is launched. 
Even under the assumption of the \texttt{MIN} IE model, our limits are comparable with the ones from a combined analysis of a sample of TeV blazars detected by HAWC~\cite{Jacobsen:2022swa}, and the ones from the spectral analysis of very-high-energy gamma rays from the blazar Mrk421~\cite{Li:2020pcn}.
Both limits are affected by severe uncertainties in the blazar jet and intracluster magnetic field 
models.
Analogous uncertainties related to the configuration of the intracluster magnetic
field can substantially weaken the limits from the search of ALP-photon induced spectral oscillations 
from the active galactic nuclei in galaxy clusters NGC 1275 and PKS 2155-304, see the discussion in~\cite{Pallathadka:2020vwu}.
Our approach offers a complementary, independent, probe of the ALP parameter space accessible by current gamma-ray 
telescopes, and extends it to higher masses progressively closing the gap up to ADMX limits~\cite{2018PhRvL.120o1301D}.

Our results depend on the better determined (at least with respect to extragalactic objects) Galactic magnetic field and rely on a large-scale signal characterization which is less dependent on single-source uncertainties. The main ALP model uncertainty is here represented by
the assumptions about the structure and redshift evolution of the magnetic field extension and strength in extragalactic sources.
Assuming either a constant source magnetic field or, instead, its redshift evolution 
induces a variation of about 50\% of the final limits.
Improvements in the determination of the major source class contributor to the 
astrophysical neutrino flux will allow us to better refine the source model for gamma-ray production 
and ALPs propagation, and to reduce the uncertainties related to the modeling of the source magnetic field.

As for background modeling uncertainties, we stress that we have assumed, at least for \tas~an optimistic instrumental threshold to point-like and extended source detection: Such a lower bound on the experimentally achievable detection threshold sets a conservative bound on the ALP-photon coupling.
The dominant uncertainty affecting the bound is instead related to the IE model. 
%\sout{We remark and caution that the IE models in Ref.~\cite{Luque:2022buq} use a particular model for a sub-threshold source population in order to match \Fermi-LAT data in the relevant energy band. Although the final prediction of the IE model's diffuse flux does not incorporate this contribution, this procedure may entail some additional source of error.} \FC{Maybe be too technical especially if the referee ask more details} 
Given the scarcity of data from TeV-bright sources, the uncertainty attributed to the unresolved Galactic population may only be reduced by increasing the sample size of detected sources via instruments like LHAASO and CTA with future observations.
We also stress that the absorption of Galactic gamma rays 
induced by conversion of these photons into ALPs, albeit already quite small, is surely overestimated, reducing our final signal strength and therefore leading
to a weaker, but conservative, bound. 

Given these uncertainties of the ``guaranteed'' Galactic astrophysical contribution to the sub-PeV measurement of Tibet AS$\gamma$ and HAWC, we checked that setting upper limits on the photon-ALP coupling $g_{a\gamma\gamma}$ for $m_a \leq 2\times10^{-7}$ eV assuming merely the ALP component itself (and no other astrophysical contribution from IE and sub-threshold sources) deteriorates the constraints by a factor of about 3 compared to the result stated in Eq.~\ref{eq:upper_limit}, still slightly stronger than CAST and HB limits~\cite{CAST:2017uph, PhysRevLett.113.191302}.
This rather conservative approach allows us to gauge what we can gain in modeling contributions to the
guaranteed astrophysical background.

In the future, besides improvements on the sub-PeV Galactic diffuse emission modeling and 
measurements, ALPs -- and more in general exotic physics -- searches will 
benefit from observations at higher latitude, where also the Galactic (diffuse and source) emission is suppressed.
Interestingly, \tas~has sensitivity also to high-latitude ($|b|>20^\circ$) photons, and,
recently, Ref.~\cite{2021A&A...653L...4N} has used the full charged cosmic-ray \tas~measurements to set an
upper limit on the diffuse gamma-ray emission for $|b|>20^\circ$.
By using these data as reported in~\cite{2021A&A...653L...4N}, we 
obtain, however, a bound which is a factor of about 3 weaker than our constraint in Eq.~\ref{eq:upper_limit}, i.e.~using the scenario of maximal IE in the sub-PeV energy range. Nonetheless, the uncertainty of the IE at higher latitudes is significantly reduced compared to the Galactic disk so that future observations by LHAASO are expected to set robust, world-leading constraints 
from high-latitude observations~\cite{LHAASO:2019qtb}.
%Nonetheless, .

\begin{figure}[t!]
\vspace{0.cm}
\includegraphics[width=0.975\columnwidth]{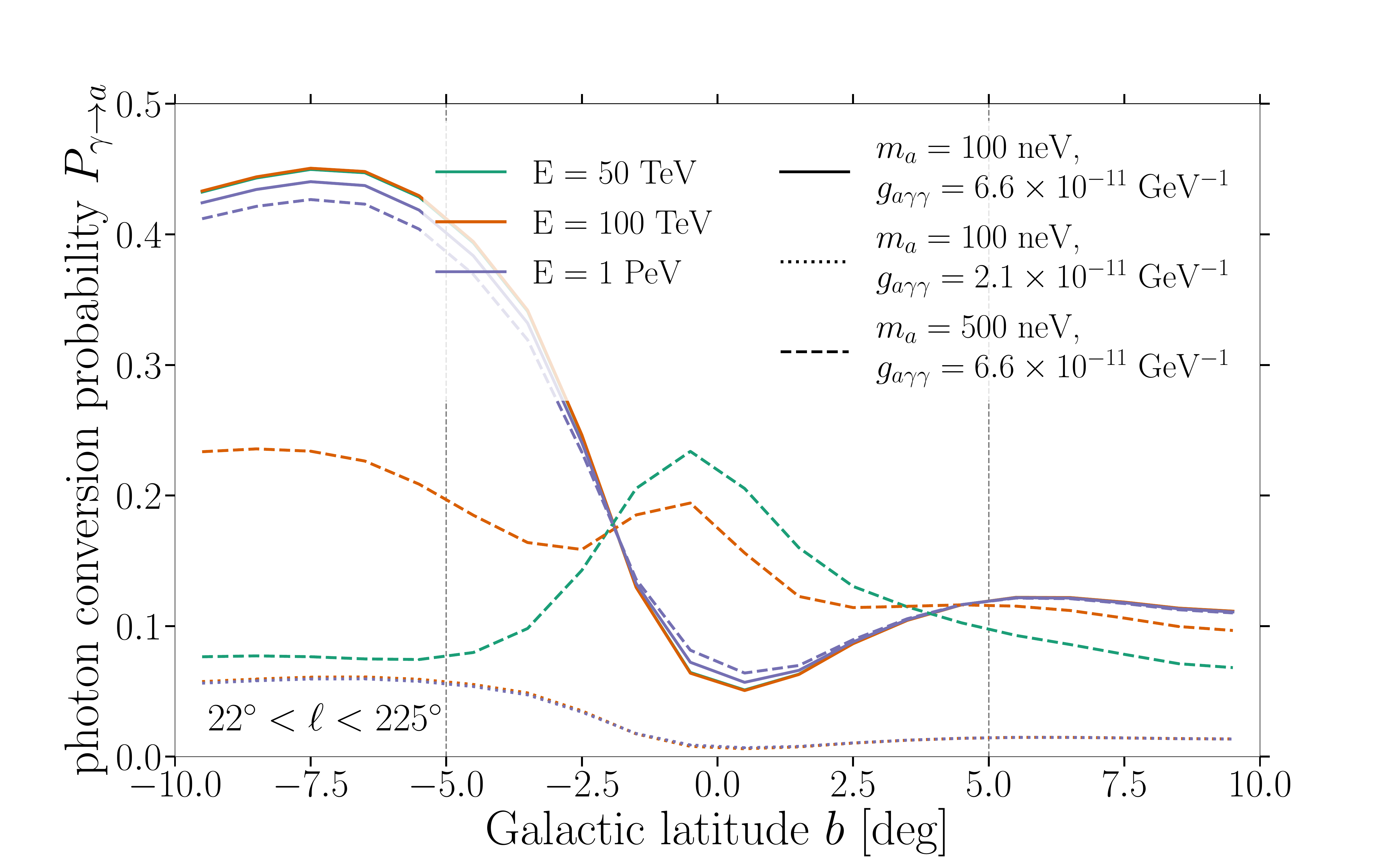}
\caption{Galactic latitude profile of the average photon-ALP conversion probability $P_{\gamma\rightarrow a}$. The average has been performed with respect to the full longitude extension and strips of $1^{\circ}$ height in Galactic latitude. We have selected three cases of ALP parameters: (1, solid) $m_a = 100$ neV and $g_{a\gamma\gamma} = 6.6\times10^{-11}$ GeV$^{-1}$ (CAST upper bound \cite{CAST:2017uph}); (2, dotted) $m_a = 100$ neV and $g_{a\gamma\gamma} = 2.1\times10^{-11}$ GeV$^{-1}$ (upper bound from this work) and (3, dashed) $m_a = 500$ neV and $g_{a\gamma\gamma} = 6.6\times10^{-11}$ GeV$^{-1}$. We show the profiles for three different energies: 50 TeV (green), 100 TeV (orange) and 1 PeV (purple). The vertical dashed lines indicate the boundaries of Tibet AS$\gamma$'s FOV in Galactic latitude. 
\label{fig:latitude_ALP}}
\end{figure}

Apart from the measurement of the diffuse flux in selected regions of interest, the Tibet AS$\gamma$ collaboration also published latitude profiles of the observed sub-PeV gamma-ray events per energy bin taking into account the entire FOV of the array, i.e.~$22^{\circ} < \ell < 225^{\circ}$, $|b| < 5^{\circ}$. However, these profiles only show the excess counts and lack a proper conversion into a gamma-ray flux. Since we lack the required information about the detector specifications and geometry, exposure as well as event selection/reconstruction efficiency, such a conversion cannot be performed on our side. As already pointed out in Ref.~\cite{2021PhRvD.104b1301E}, the Tibet AS$\gamma$ latitude profiles contain valuable information and constraining power for multiple classes of ``New Physics''. 
We provide an illustrative plot of the latitude profile of the average photon-ALP conversion probability $P_{\gamma\rightarrow a}$ in the full Tibet AS$\gamma$ FOV in Fig.~\ref{fig:latitude_ALP}, assuming $m_a = 100$ neV and $g_{a\gamma\gamma} = 6.6\times10^{-11}$ GeV$^{-1}$ (CAST upper limit \cite{CAST:2017uph}) and the Jansson \& Farrar Galactic magnetic field model as a solid line.
On one side, the latitude profile of the diffuse ALP signal therefore strongly depends on the morphology of the Milky Way's Galactic magnetic field, attaining its maximal values within the Galactic disk (regions with highest $P_{\gamma\rightarrow a}$ in Fig.~\ref{fig:latitude_ALP}). On the other side, the significance of the effect is also determined by the value of the ALP-photon coupling $g_{a\gamma\gamma}$ (see dotted lines in Fig.~\ref{fig:latitude_ALP} illustrating the scenario at the upper limit stated in Eq.~\ref{eq:upper_limit}) and the mass of the ALP (dashed lines in the same figure for $m_a = 500$ neV). The latter choice of ALP mass was made to exemplify the behavior in the regime where we lose sensitivity to the exotic extragalactic ALP signal (see Fig.~\ref{fig:ULIM_ALPs}). In this mass range the oscillatory nature of the ALP-photon conversion becomes relevant. At 1 PeV this effect is still suppressed while it is present at 50 TeV and 100 TeV altering the general trend seen in all other plotted scenarios.

%\sout{As concerns the latitude profile of the diffuse ALP signal \CE{-- for which we provide the latitude profile of the average photon-ALP conversion probability $P_{\gamma\rightarrow a}$ in the full Tibet AS$\gamma$ FOV in Fig.~\ref{fig:latitude_ALP} assuming $m_a = 100$ neV and $g_{a\gamma\gamma} = 5\times10^{-11}$ GeV$^{-1}$ as well as the Jansson \& Farrar Galactic magnetic field model --} it is strongly depending on the morphology of the Milky Way's Galactic magnetic field, which exhibits particular features and which attains its maximal values within the Galactic disk \CE{(regions with highest $P_{\gamma\rightarrow a}$ in Fig.~\ref{fig:latitude_ALP})}}.
Hence, an angular analysis of the expected ALP signal may be even more constraining than the analysis performed in this work. Besides its applicability to Tibet AS$\gamma$ data, the ALP-induced anisotropy of gamma rays due to the particular structure of the Galactic magnetic field is a universal feature that may be detected by air shower arrays similar to Tibet AS$\gamma$ and HAWC or Imaging Atmospheric Cherenkov telescopes like H.E.S.S.~or next-generation instruments as, e.g.,~CTA that are capable of probing astrophysically produced gamma rays above 10 TeV (as already pointed our in, e.g., Ref.~\cite{Simet:2007sa}). Moreover, the anisotropies' morphology is energy-dependent in case of massive ALPs as demonstrated by the dashed lines in Fig.~\ref{fig:latitude_ALP} offering a means to constrain $g_{a\gamma\gamma}$ and $m_a$ at the same time.

At these energies, such a search can already be conducted at the level of event reconstruction and rejection of hadronic extensive air showers: While a ``gamma-cleaned'' sample of events, i.e.~air showers classified as genuine primary gamma rays, exhibits an ALP-induced anisotropy, the corresponding sample of rejected air showers triggered by hadronic primary cosmic rays, should not feature such anisotropy. The large survey regions of the planned extragalactic and Galactic plane survey of CTA (see Ref.~\cite{CTAConsortium:2017dvg} for the envisioned survey regions and observation schemes) provide a suitable basis to perform such a search for ALPs. In order to mitigate the uncertainty of the spatial morphology of Galactic gamma-ray contribution at these energies, a combination of several independent high- and low-latitude ROIs that exhibit the sought-after anisotropy may further enhance the sensitivity to ALP interactions with photons.

\medskip
%%%%%%%%%%%%%%%%%%%%%%%%%%%%%%%%%%%%%%
\textbf{Acknowledgments.} 
We warmly thank Pasquale D.~Serpico and Alessandro Mirizzi for their
careful reading of the manuscript and inspiring comments.
We also acknowledge helpful discussions and result comparison with Pierluca Carenza and Leonardo Mastrototaro. We thank the anonymous referee for their valuable comments on our work.
We thank Ciaran O'Hare for providing the publicly available collection of current ALP upper limits and the plot generating software~\cite{AxionLimits}. We are likewise grateful to Manuel Meyer, James Davies and Julian Kuhlmann for creating and maintaining the python package \texttt{gammaALPs} \cite{meyer_manuel_2021_6344566} that we have made extensively use of. 
We acknowledge support by the ``Agence Nationale de la Recherche'', grant n. ANR-19-CE31-0005-01 (PI: F. Calore). 
%
%%%%%%%%%%%%%%%%%%%%%%%%%%%%%%%%%%%%%%

\medskip
%%%%%%%%%%%%%%%%%%%%%%%%%%%%%%%%%%%%%%
\textbf{Data Availability.}
The digitized data and \textsc{python} scripts utilized to produce the results of this work are publicly available under: \url{https://github.com/ceckner/subPeVALPs}.
%
%%%%%%%%%%%%%%%%%%%%%%%%%%%%%%%%%%%%%%

\bibliography{ALPs_from_SFG.bib}

%\clearpage 

\appendix

%%%%%%%%%%%%%%%%%%%%%%%%%%%%%%%
\section{Modeling the gamma-ray contribution from a synthetic Galactic source population}
\label{app:subthresh_population}
%%%%%%%%%%%%%%%%%%%%%%%%%%%%%%%

The model developed in Ref.~\cite{2021arXiv210714584V} is based on the number density $\rho$ and luminosity function $\mathcal{L}$ of a synthetic Galactic population according to:
\begin{equation}
\label{eq:source_density}
    \frac{\diff N}{\diff^3r\,\diff L_{\mathrm{TeV}}} = \rho\!\left(\bm{r}\right) \times \mathcal{L}(L_{\mathrm{TeV}})\mathrm{,}
\end{equation}
where the axially symmetric $\rho\!\left(\bm{r}\right)$ is defined as the product of the radial pulsar distribution reported in Ref.~\cite{Lorimer:2006qs} and an exponential function
\begin{equation}
\label{eq:spatial_source_density}
    \rho\!\left(\bm{r}\right) \sim \left(\frac{r}{r_{\odot}}\right)^B\exp{\left[-C\left(\frac{r- r_{\odot}}{r_{\odot}}\right)\right]} \exp{\left(-\frac{|z|}{H}\right)}\mathrm{,}
\end{equation}
with $r_{\odot} = 8.5$ kpc, $B = 1.9$, $C = 5.0$ and $H = 0.2$ kpc. The values of the parameters $B$ and $C$ correspond to the best-fit for Model C in Ref.~\cite{Lorimer:2006qs}, which is coined therein as the optimal model describing the distribution of pulsars in the Galactic disk. 
%We normalize $\rho\!\left(\bm{r}\right)$ to unity when integrated over the full volume of the Milky Way.

The luminosity function $\mathcal{L}$ characterizes the intrinsic luminosity distribution of all objects constituting the synthetic TeV-bright population. It reads:
\begin{equation}
\label{eq:luminosity_function}
    \mathcal{L}(L_{\mathrm{TeV}}) = \frac{R\tau(\alpha - 1)}{L_{\mathrm{TeV}, \mathrm{max}}}\left(\frac{L_{\mathrm{TeV}}}{L_{\mathrm{TeV}, \mathrm{max}}}\right)^{-\alpha}
\end{equation}
Here, $L_{\mathrm{TeV}}$ refers to the gamma-ray luminosity of an individual object of the population in the energy band from 1 to 100 TeV. The free parameters of this expression have been derived in Ref.~\cite{Cataldo:2020qla} from the H.E.S.S.~Galactic plane survey results and its resulting source catalog \cite{HESS:2018pbp}. We adopt the nominal values as stated by the authors, i.e.~$L_{\mathrm{TeV}, \mathrm{max}} = 4.9\times10^{35}\;\mathrm{erg}\,\mathrm{s}^{-1}$, $\tau = 1.8\times10^3$ yr and $\alpha = 1.5$. Besides, this ``reference scenario'' the authors of Ref.~\cite{Cataldo:2020qla} provide a more aggressive set of parameters that generates slightly larger fluxes. However, we will not make use of the latter for the sake of remaining conservative in our final ALPs limits.

To translate the intrinsic luminosity of each source into a gamma-ray flux $\Phi$, we have to assume an average gamma-ray spectrum $\varphi\!\left(E\right)$. To this end, we select a power law with exponential cutoff
\begin{equation}
\label{eq:avg_spectrum}
\varphi\!\left(E\right) = K_0 \left(\frac{E}{1\;\mathrm{TeV}}\right)^{-\beta}\exp{\left(-\frac{E}{E_{\mathrm{c}}}\right)}\mathrm{,}
\end{equation}
where the value of $K_0$ follows from the requirement that Eq.~\ref{eq:avg_spectrum} is normalized to one when integrated from 1 TeV to 100 TeV. The spectral parameters $\beta$ and $E_{\mathrm{c}}$ are free parameters of the model.

As a last step, we obtain the cumulative flux $\Phi^{\mathrm{sTH}}$ of all TeV-bright sources below a detection threshold $S_{\mathrm{TH}}$ via
\begin{equation}
\label{eq:cumulative_flux}
\Phi^{\mathrm{sTH}}\!\left(E\right) = \varphi\!\left(E\right) \intop_{0}^{S_{\mathrm{TH}}} \Phi_{\mathrm{TeV}}\frac{\diff N}{\diff \Phi_{\mathrm{TeV}}}\,\diff\Phi_{\mathrm{TeV}}\mathrm{,}
\end{equation}
where the differential number of sources per unit flux $\Phi_{\mathrm{TeV}}$\footnote{Note that the flux variable $\Phi_{\mathrm{TeV}}$ is again valid for the energy band from 1 to 100 TeV.} is directly related to Eq.~\ref{eq:source_density} under the change of variable $L_{\mathrm{TeV}} = 4\pi d^2\Phi_{\mathrm{TeV}}\langle E\rangle$ and integrating out the spatial dependence by substituting the boundaries of the region of interest for a particular instrument and data set (see Ref.~\cite{Cataldo:2020qla} for further details). In this framework, $d$ is the distance of an object to the Earth, while $\langle E\rangle$ refers to the average photon energy of a source in the population given the assumed average spectrum $\varphi(E)$ and energy band from 1 to 100 TeV, i.e.:
\begin{equation}
\label{eq:cumulative_flux}
\langle E\rangle = \frac{\intop_{1\;\mathrm{TeV}}^{100\;\mathrm{TeV}} E\varphi(E)\,\diff E}{\intop_{1\;\mathrm{TeV}}^{100\;\mathrm{TeV}} \varphi(E)\,\diff E}\mathrm{.}
\end{equation}

%%%%%%%%%%%%%%%%%%%%%%%%%%%%%%%
\section{Fundamentals of photon-ALP mixing}
\label{app:gamma_ALP_mixing}
%%%%%%%%%%%%%%%%%%%%%%%%%%%%%%%

The physics of mixing between photon and ALP states, i.e.~the Primakoff process, follows directly from the Lagrangian in Eq.~\ref{eq:ALP_lagrangian}. In what follows, we do not aim at giving a precise re-iteration of the formalism that has been developed and refined over the last decade(s) but we emphasize the basic equations and ingredients that are necessary to calculate the probability that photon states undergo a conversion into ALPs and vice versa. More complete and rigorous treatments of both the physical and the mathematical aspects of ALP propagation and conversion can be found -- without the intent of providing an exhaustive list -- in Refs.~\cite{Raffelt:1987im, DeAngelis:2007dqd, Mirizzi:2009aj, DeAngelis:2011id, Meyer:2014epa, Kartavtsev:2016doq}.

The Primakoff process requires the existence of an external magnetic field $\bm{B}$ with a non-vanishing component $\bm{B}_{\perp}$ transversal to the propagation direction of an initial photon or ALP state. Without loss of generality, we assume that the initial state with energy $E$ propagates in $\bm{\hat{z}}$-direction while the magnetic field is described according to $\bm{B}_{\perp} = B\left(\cos\theta, \sin\theta, 0\right)^T$ such that $\cos\theta$ is the polar angle between the direction of the transversal magnetic field $\bm{B}_{\perp}$ and the $\bm{\hat{x}}$-direction of the transversal plane spanned by $\bm{\hat{x}}$ and $\bm{\hat{y}}$. Let $A_x$ and $A_y$ denote the respective photon polarization states. 

In this setting, the evolution/propagation in $\bm{\hat{z}}$-direction of a pure photon-ALP state is given by \cite{Raffelt:1987im, Kartavtsev:2016doq}
\begin{equation}
\label{eq:ALP-photon-prop-pure}
i\frac{\diff\mathcal{A}}{\diff z} = \left(\mathcal{H}_{\mathrm{dis}} - \frac{i}{2} \mathcal{H}_{\mathrm{abs}}\right) \mathcal{A}\mathrm{,}
\end{equation}
where $\mathcal{A} = \left(A_{\perp}, A_{\parallel}, a\right)^T$ defines the three-component wave function that contains the ALP state $a$ and the two photon polarization states in the transversal plane denoted by $A_{\perp}$ (perpendicular to the direction of $\bm{B}_{\perp}$) and $A_{\parallel}$ (parallel to the direction of $\bm{B}_{\perp}$). The photon polarization states are a linear combination of  $A_x$ and $A_y$ accordingly weighted by the polar angle $\theta$. Note that only polarization states parallel to the transversal magnetic field component can convert to ALP states.

The Hamiltonian $\mathcal{H}_{\mathrm{abs}}$ characterizes the losses due to photon absorption in the particular environment under study, for instance, Galactic absorption or absorption on the CMB and EBL. Absorption processes may also occur for ALPs but they scale quadratically with the coupling strength $g_{a\gamma\gamma}$ and are thus highly suppressed. Therefore, this component can be expressed as
\begin{equation}
\label{eq:absoprtion_matrix}
\mathcal{H}_{\mathrm{abs}} = \left(\begin{array}{ccc}
\Gamma & 0 & 0\\
0 & \Gamma & 0\\
0 & 0 & 0
\end{array}\right)
\end{equation}
under moderate assumptions about the properties of the environments \cite{Kartavtsev:2016doq}, where $\Gamma$ quantifies the (energy-dependent) photon-photon absorption strength whose analytic expression can be found in Ref.~\cite{Mirizzi:2009aj}.

The other Hamiltonian $\mathcal{H}_{\mathrm{dis}}$ takes into account dispersion effects in the photon-ALP state, i.e.~processes inducing a conversion of $A_{\parallel}$ and $a$. In the discussed geometrical framework, it reads (neglecting the contribution due to Farraday rotation) \cite{Raffelt:1987im, Mirizzi:2009aj}, 
\begin{equation}
\label{eq:mixing_matrix}
\mathcal{H}_{\mathrm{dis}} = \left(\begin{array}{ccc}
\Delta_{\perp} & 0 & 0\\
0 & \Delta_{\parallel} & \Delta_{a\gamma}\\
0 & \Delta_{a\gamma} & \Delta_{a}
\end{array}\right)\mathrm{,}
\end{equation}
where the individual matrix elements are defined as follows \cite{Mirizzi:2009aj, Kartavtsev:2016doq}:
\begin{align}
\label{eq:mixing_components}
\Delta_{\perp} &= \Delta_{\mathrm{pl}} + 2\Delta_{\mathrm{B}} + \Delta_{\gamma\gamma}\\
\Delta_{\parallel} &= \Delta_{\mathrm{pl}} + \frac{7}{2}\Delta_{\mathrm{B}} + \Delta_{\gamma\gamma}\\
\Delta_a &= -\frac{m^2_a}{2E}\\
\Delta_{a\gamma} &= \frac{g_{a\gamma\gamma}}{2}B\mathrm{.}
\end{align}
The dispersion matrix elements concerning the perpendicular and parallel photon polarization states are the result of different phenomena, i.e.~refraction on the electron plasma with electron density $n_e$ in the environment under study -- $\Delta_{\mathrm{pl}} = - 2\pi\alpha n_e / (m_eE)$; refraction on the magnetic field with energy density $\rho_B = 1/2|\bm{B}|^2$ -- $\Delta_{\mathrm{B}} = 24\alpha^2\rho_B/(135m_e^4)\sin^2\!\theta E$ as well as photon-photon-dispersion on radiation fields like the CMB, in which case we find -- $\Delta_{\gamma\gamma} \approx 44\alpha^2\rho_{\mathrm{CMB}} E / (135m_e^2)$. Further radiation fields can in principle contribute to $\Delta_{\gamma\gamma}$. For the computation of the dispersive part of the refraction index associated to these contributions, we use the prescription in Ref.~\cite{Dobrynina:2014qba}.

In reality and for all practical purposes, working with pure states is an oversimplification of the problem. A better description is to transform Eq.~\ref{eq:ALP-photon-prop-pure} into its analog with respect to density matrices. The density matrix of the photon-ALP system is constructed via $\bm{\rho} = \mathcal{A}\bigotimes\mathcal{A}^{\dagger}$. In this case, the evolution of the density matrix is described by \cite{Mirizzi:2009aj, Kartavtsev:2016doq} 
\begin{equation}
\label{eq:ALP-photon-prop-density}
i\frac{\diff\bm{\rho}}{\diff z} = \left[ \mathcal{H}_{\mathrm{dis}}, \bm{\rho}\right] - \frac{i}{2} \{\mathcal{H}_{\mathrm{abs}}, \bm{\rho}\}\mathrm{.}
\end{equation}

%%%%%%%%%%%%%%%%%%%%%%%%%%%%%%%
\section{Tabulated flux values at different exclusion levels}
\label{app:likelihood_values}
%%%%%%%%%%%%%%%%%%%%%%%%%%%%%%%

In this section, we present in Tab.~\ref{tab:ALPflux_vs_E} the results of our statistical analysis in terms of the maximally allowed ALP flux (for ALP masses $m_a\lesssim 2\times10^{-7}$ eV) over the energy range of interest (adhering to the binning scheme employed to the HAWC flux and as stated by the Tibet AS$\gamma$ collaboration) for different exclusion levels.

\clearpage
\thispagestyle{empty}
%\vspace*{20cm}

\begin{sidewaystable*}
\begin{centering}
{\renewcommand{\arraystretch}{2.0}
\begin{tabular}{|c c|c|c|c|c|c|c|c|c|c|c|}
\hline 
\hline
\multirow{2}{*}{$E_{\mathrm{min}}\,\left[\mathrm{TeV}\right]$} & \multirow{2}{*}{$E_{\mathrm{max}}\,\left[\mathrm{TeV}\right]$} & \multicolumn{5}{c|}{IE \texttt{MAX}$:\ensuremath{\Phi_{X\mathrm{C.L.}}^{\mathrm{ALP}}}\;\left[\mathrm{GeV}^{1.7}\mathrm{cm}^{-2}\mathrm{s}^{-1}\mathrm{sr}^{-1}\right]$} & \multicolumn{5}{c|}{IE \texttt{MIN}$:\ensuremath{\Phi_{X\mathrm{C.L.}}^{\mathrm{ALP}}}\;\left[\mathrm{GeV}^{1.7}\mathrm{cm}^{-2}\mathrm{s}^{-1}\mathrm{sr}^{-1}\right]$}\tabularnewline
 &  & \multicolumn{1}{c}{$68.3\%$} & \multicolumn{1}{c}{$90\%$} & \multicolumn{1}{c}{$95\%$} & \multicolumn{1}{c}{$99\%$} & $99.9\%$ & \multicolumn{1}{c}{$68.3\%$} & \multicolumn{1}{c}{$90\%$} & \multicolumn{1}{c}{$95\%$} & \multicolumn{1}{c}{$99\%$} & $99.9\%$\tabularnewline
\hline 
10.0 & 15.9 & $2.94\times10^{-6}$ & $5.32\times10^{-6}$ & $\ensuremath{7.20\times10^{-6}}$ & $\ensuremath{1.04\times10^{-5}}$ & $\ensuremath{1.34\times10^{-5}}$ & $1.22\times10^{-5}$ & $\ensuremath{1.49\times10^{-5}}$ & $\ensuremath{1.72\times10^{-5}}$ & $\ensuremath{2.17\times10^{-5}}$ & $\ensuremath{2.61\times10^{-5}}$\tabularnewline
15.9 & 25.1 & $4.96\times10^{-6}$ & $9.00\times10^{-6}$ & $\ensuremath{1.22\times10^{-5}}$ & $\ensuremath{1.79\times10^{-5}}$ & $\ensuremath{2.30\times10^{-5}}$ & $2.09\times10^{-5}$ & $\ensuremath{2.57\times10^{-5}}$ & $\ensuremath{2.97\times10^{-5}}$ & $\ensuremath{3.76\times10^{-5}}$ & $\ensuremath{4.52\times10^{-5}}$\tabularnewline
25.1 & 39.8 & $6.50\times10^{-6}$ & $1.18\times10^{-5}$ & $\ensuremath{1.62\times10^{-5}}$ & $\ensuremath{2.38\times10^{-5}}$ & $\ensuremath{3.06\times10^{-5}}$ & $2.78\times10^{-5}$ & $\ensuremath{3.43\times10^{-5}}$ & $\ensuremath{3.97\times10^{-5}}$ & $\ensuremath{5.04\times10^{-5}}$ & $\ensuremath{6.08\times10^{-5}}$\tabularnewline
39.8 & 63.1 & $6.38\times10^{-6}$ & $1.16\times10^{-5}$ & $\ensuremath{1.60\times10^{-5}}$ & $\ensuremath{2.38\times10^{-5}}$ & $\ensuremath{3.07\times10^{-5}}$ & $2.79\times10^{-5}$ & $\ensuremath{3.44\times10^{-5}}$ & $\ensuremath{3.99\times10^{-5}}$ & $\ensuremath{5.08\times10^{-5}}$ & $\ensuremath{6.14\times10^{-5}}$\tabularnewline
63.1 & 100.0 & $4.96\times10^{-6}$ & $9.06\times10^{-6}$ & $\ensuremath{1.25\times10^{-5}}$ & $\ensuremath{1.87\times10^{-5}}$ & $\ensuremath{2.43\times10^{-5}}$ & $2.20\times10^{-5}$ & $\ensuremath{2.73\times10^{-5}}$ & $\ensuremath{1.25\times10^{-5}}$ & $\ensuremath{4.04\times10^{-5}}$ & $\ensuremath{4.88\times10^{-5}}$\tabularnewline
\hline 
100.0 & 158.0 & $3.73\times10^{-6}$ & $6.82\times10^{-6}$ & $\ensuremath{9.43\times10^{-6}}$ & $\ensuremath{1.41\times10^{-5}}$ & $\ensuremath{1.83\times10^{-5}}$ & $1.66\times10^{-5}$ & $\ensuremath{2.06\times10^{-5}}$ & $\ensuremath{2.39\times10^{-5}}$ & $\ensuremath{3.03\times10^{-5}}$ & $\ensuremath{3.65\times10^{-5}}$\tabularnewline
158.0 & 398.0 & $1.74\times10^{-6}$ & $3.20\times10^{-6}$ & $\ensuremath{4.44\times10^{-6}}$ & $\ensuremath{6.67\times10^{-6}}$ & $\ensuremath{8.68\times10^{-6}}$ & $7.85\times10^{-6}$ & $\ensuremath{9.75\times10^{-6}}$ & $\ensuremath{1.13\times10^{-5}}$ & $\ensuremath{1.44\times10^{-5}}$ & $\ensuremath{1.74\times10^{-5}}$\tabularnewline
398.0 & 1000.0 & $6.78\times10^{-6}$ & $1.25\times10^{-6}$ & $\ensuremath{1.75\times10^{-6}}$ & $\ensuremath{2.67\times10^{-6}}$ & $\ensuremath{3.50\times10^{-6}}$ & $3.16\times10^{-6}$ & $\ensuremath{3.95\times10^{-6}}$ & $\ensuremath{4.59\times10^{-6}}$ & $\ensuremath{5.87\times10^{-6}}$ & $\ensuremath{7.10\times10^{-6}}$\tabularnewline
\hline 
\hline
\end{tabular}
}
\par\end{centering}
\caption{Maximally allowed ALP-induced gamma-ray flux for $m_a\lesssim 2\times10^{-7}$ eV that contributes to the Galactic diffuse measurement of HAWC and Tibet AS$\gamma$ at the 68.3\%, 90\%, 95\%, 99\% and 99.9\% C.L. The first five rows indicate the flux associated to the energy range of the HAWC diffuse measurement (ROI: $43^\circ < l < 73^\circ$ and $|b| < 4^\circ$) while the last three rows state the corresponding values for Tibet AS$\gamma$ (ROI: $25^\circ < l < 100^\circ$ and $|b| < 5^\circ$). The first and second columns list the lower and upper boundary of the considered energy bins. The adjacent block states the allowed flux level assuming our model of maximal IE whereas the last block contains the respective values for the minimal IE model. The contribution from sub-threshold sources is always included in the astrophysical model part. These values furthermore reflect the assumption of the fiducial prescription of the star-formation rate density evolution of the universe, as well as scenario \emph{(i)} of the magnetic field strength evolution in extragalactic neutrino-generating sources. \label{tab:ALPflux_vs_E}}
\end{sidewaystable*}

\end{document}